\documentclass[lettersize,journal]{IEEEtran}
\usepackage{amsmath,amsfonts}
\usepackage{algorithmic}
\usepackage{algorithm}
\usepackage{array}
\usepackage[caption=false,font=normalsize,labelfont=sf,textfont=sf]{subfig}
\usepackage{textcomp}
\usepackage{stfloats}
\usepackage{url}
\usepackage{verbatim}
\usepackage{graphicx}
\usepackage{cite}
\usepackage{color}
\usepackage{amssymb}
\usepackage{bm}
\usepackage{booktabs}
\usepackage{tabularx}
\begin{document}

\title{Secure Relay Low‑Altitude Networks via Hybrid Fixed‑Position and Rotatable Antenna Arrays}

\author{\textcolor{black}{Maolin Li, Feng Shu, Minghao Chen, Riqing Chen, Wei Feng, Wei Gao, Qi Zhang, Liang Yang, Cunhua Pan}
	\thanks{This work was supported in part by the National Natural Science Foundation of China under Grant U22A2002, and  by the Hainan Province Science and Technology Special Fund under Grant ZDYF2024GXJS292; in part by the Scientific Research Fund Project of Hainan University under Grant KYQD(ZR)-21008; in part by the Collaborative Innovation Center of Information Technology, Hainan University, under Grant XTCX2022XXC07; in part by the National Key Research and Development Program of China under Grant 2023YFF0612900. (Corresponding authors: \textcolor{black}{Minghao Chen and} Feng Shu.) }
		\thanks{Maolin Li, \textcolor{black}{Minghao Chen}, and Qi Zhang are with the School
		of Information and Communication Engineering, Hainan University, Haikou,
		570228, China. (e-mail: limaolin0302@163.com; \textcolor{black}{996739@hainanu.edu.cn}; hdzhangqi0509@163.com).}
				\thanks{Riqing Chen is with the Digital Fujian Institute of Agricultural Big Data, Fujian Agriculture and Forestry University, Fujian, 350002, China, and also with the Fujian Key Lab of Agricultural IOT Applications, Sanming University, Fujian, 365004, China. (e-mail: riqing.chen@fjsmu.edu.cn).}	
				\thanks{\textcolor{black}{Wei Feng is with the Department of Electronic Engineering, State Key Laboratory of Space Network and Communications, Tsinghua University, Beijing 100084, China (e-mail: fengwei@tsinghua.edu.cn).}}
				\thanks{Wei Gao is with the School of Information and Communication Engineering, Hainan University, Haikou 570228, China, also with CSG Digital Power Grid Research Institute Company Ltd., Guangzhou 510555, China, and also with China Energy Research Society, Beijing 100045, China. (e-mail: gaowei@hainanu.edu.cn).}
			\thanks{Feng Shu is with the School of Information and Communication Engineering, Hainan University, Haikou 570228, China, and also with the School of Electronic and Optical Engineering, Nanjing University of Science and Technology, Nanjing 210094, China. (e-mail: shufeng0101@163.com).}	
			\thanks{\textcolor{black}{Liang Yang is with the College of Computer Science and Electronic Engineering, Hunan University, Changsha 410082, China, and also with School of Information Engineering, Changsha Medical University, Changsha 410219, China. (e-mail: liangy@hnu.edu.cn).} }
	\thanks{Cunhua Pan is with the National Mobile Communications Research Laboratory, Southeast University, China. (e-mail: cpan@seu.edu.cn).}	
}

\markboth{Journal of \LaTeX\ Class Files,~Vol.~14, No.~8, August~2021}%
{Shell \MakeLowercase{\textit{et al.}}: A Sample Article Using IEEEtran.cls for IEEE Journals}


\maketitle

\begin{abstract}
In this paper, a relay network with hybrid fixed‑position and rotatable antenna arrays is proposed. The deployment of rotatable arrays in conventional relay networks is considered to provide more secure communications for low‑altitude economy applications. Specifically, both the base station and the relay station are equipped with fixed‑position antenna arrays and rotatable arrays to serve ground users and aerial users, respectively. To address the challenge of multi‑user interference, a low‑cost reconfigurable intelligent surface is exploited as a candidate path. Accordingly, under constraints on transmit power, user quality of service, rotatable range, and path selection, the objective is to maximize the worst‑case secrecy rate (SR) through joint beamforming, power allocation, and rotatable antenna orientation design. First, the SR performance in the \textcolor{black}{single ground–aerial user pair scenario} is investigated, and a step‑by‑step leakage‑based scheme is proposed. Then, the general multi‑user scenario is studied, and a Distributional Soft Actor-Critic with Three refinements (DSAC-T)‑based learning scheme, which supports hybrid discrete and continuous actions, is proposed to maximize the worst‑case SR. Simulation results validate the effectiveness of the proposed schemes. The proposed schemes achieve approximately a twofold improvement in SR performance compared to isotropic antennas. \textcolor{black}{The proposed system can achieve approximately 99.7\% power saving, 55\% antenna saving, and can serve more users.}
\end{abstract}


\section{Introduction}
\IEEEPARstart{W}{ith} the surge in user numbers and density, sixth‑generation (6G) wireless communication technologies have been extensively investigated. To achieve high channel capacity and spectral efficiency to meet 6G requirements, ultra‑massive multiple input multiple-output (MIMO) technology has been studied, offering high array gain and spatial multiplexing gain for wireless networks~\cite{Jiang2025a}. However, it is impractical due to the expensive hardware and implementation costs. Moreover, the spatial degrees-of-freedom (DoFs) cannot be fully exploited because of fixed positions and orientations. To address this challenge, reconfigurable intelligent surface (RIS)~\cite{Huang2022}, fluid antenna~\cite{Wong2021,Li2026c,Jiang2025}, movable antenna (MA)~\cite{Zhu2026,Li2025c,Li2026b}, and rotatable antenna (RA) have been explored~\cite{Zheng2026b}.

By reconfiguring the wireless propagation environment, RIS offers a low-cost, energy-efficient, and easy-to-deploy solution for performance enhancement in 6G networks~\cite{Zhang2025a}. Its fundamental value lies in shifting the paradigm from traditional channel adaptation to channel shaping, thereby enabling proactive control over the electromagnetic space. With the maturation of hardware technologies and the advancement of optimization algorithms, RIS is poised to play a pivotal role in enhancing network performance, expanding application scenarios, and reducing energy consumption, serving as a core component in building intelligent, efficient, and green future wireless networks~\cite{ Zhang2025b}. Generally, RIS architectures can be categorized into three types: passive~\cite{Raeisi2024,Cao2024}, active~\cite{Peng2024}, and hybrid active-passive.
In contrast, active RIS integrates active amplifying circuits, such as low-noise amplifiers and power amplifiers, enabling it to regenerate and amplify reflected signals. It supports independent amplitude and phase control, facilitating complex beamforming designs. Thanks to its signal reception capabilities, active RIS offers high channel estimation accuracy with reduced pilot overhead. Several key RIS technologies have been investigated, including simultaneously transmitting and reflecting RIS~\cite{Li2024}, beyond-diagonal RIS~\cite{Zhou2024}, intelligent omni-surfaces~\cite{Zhang2022}, and flexible RISs~\cite{Yang2026}.

To enhance communication reliability, spectral efficiency, and security, cooperative communication between RISs and relays has been investigated. Relays can be deployed in complex urban scenarios to alleviate issues such as blockage, co‑channel interference, and path loss~\cite{Zhou2018,Li2025e}. However, under dense device deployment, the achievable performance of conventional relay networks is limited by interference. As a low‑cost solution, RISs can be flexibly installed on infrastructure surfaces such as building facades, thereby providing enhanced cascaded links for relay networks. The amplitude and phase shifts of active RISs can be optimized to reduce accumulated co‑channel interference. Furthermore, the trade‑off among security, energy efficiency, and spatial multiplexing has been investigated through joint optimization of the relay and RIS. In~\cite{Ji2026}, a dual‑RIS‑assisted decode‑and‑forward relay system was studied. An adaptive RIS reflecting element allocation algorithm was proposed to reduce the overall system interference while optimizing resource allocation and improving system capacity. The achievable rate performance of single‑RIS and multi‑RIS assisted decode‑and‑forward relay systems in near‑field communications was investigated in~\cite{Qian2026,Wang2022a}. It is verified that, in near‑field scenarios, deploying multiple RISs in close proximity to the relay can enhance transmission performance and outperform the single‑RIS counterpart. In~\cite{Liu2023}, the secrecy rate (SR) maximization problem was considered in the presence of an untrusted relay. An iterative optimization algorithm was proposed to jointly optimize the RIS phase shift matrix and the beamforming vectors of the multi‑antenna access point and relay, achieving superior performance over other baseline schemes.

However, conventional base stations and relays are equipped with fixed‑position and fixed‑orientation antennas, making it difficult to cope with dynamically varying wireless channel conditions. In practice, due to the cost constraints of RF chains and antennas, the achievable performance of the system is limited. To address these challenges, equipping base stations and relays with MA arrays has been investigated, where the positions of the MAs can be flexibly adjusted to improve channel quality and enhance system performance. In~\cite{Li2025d}, a relay equipped with MAs was considered, where the antenna positions and beamforming in two phases were optimized to maximize the achievable transmission rate. By employing projected gradient ascent and iterative optimization algorithms, higher network performance was achieved compared to the fixed‑position antenna relay system. To reduce latency, a delay minimization scheme was formulated in~\cite{Xiu2026} under user quality‑of‑service constraints and solved using a penalty dual‑decomposition framework, achieving higher efficiency compared to conventional fixed‑position antenna systems. The above works focus on the case where MAs are deployed only at a single node. When both the transmitting and receiving ends are equipped with MAs~\cite{Zhu2024a}, additional challenges emerge regarding channel estimation~\cite{Zhu2025, Xiao2024c}, delay~\cite{Wang2026,Li2026}, and computational complexity~\cite{Zhu2024b}. Furthermore, six‑dimensional MAs have been investigated, which allow the three‑dimensional positions and rotations of distributed arrays to be optimized to maximize the achievable transmission rate. Both continuous and discrete coordinates have been considered~\cite{Shao2025,Shao2025a}, which incur relatively high algorithmic complexity.

To further exploit the available spatial DoFs, RA architectures have been investigated. Directional antennas are employed, whose beam directions can be steered through mechanical or electronic drives without requiring three‑dimensional movement of the antennas or arrays. This reduces design complexity and offers good scalability~\cite{Zheng2026}. To reduce hardware complexity, a cross‑linked RA structure was considered in~\cite{Zheng2026b}. By jointly optimizing the receive beamforming and rotation angles at the base station, the sum rate was maximized, achieving significant performance gains over the conventional fixed-direction antennas. Under co‑channel spectrum sharing, a RA‑enabled multiple‑input single‑output system was studied in~\cite{Peng2026}, which verified that combining RAs with conventional beamforming can significantly improve the achievable transmission rate. In~\cite{Zhang2026}, a UAV mmWave Massive MIMO network equipped with RA arrays was considered. It was shown that the RA arrays enable the channel vectors of different users to become asymptotically orthogonal, achieving higher transmission efficiency compared to fixed antenna orientations. Flexible antenna architectures have been studied and shown to achieve superior performance compared to fixed-position antennas. However, research on relay networks equipped with RAs is still in its infancy, and most existing works focus on single‑node antenna architecture improvements at either the transmitter or the relay. In the pursuit of fully exploiting spatial DoFs, the trade‑off between cost and system performance remains a critical challenge. \textcolor{black}{Hybrid configurations that integrate RAs with conventional fixed antenna arrays have been identified as a promising deployment architecture~\cite{Zheng2026d, Zheng2025}. In such a configuration, fixed antennas can maintain traditional ground coverage, while RAs provide additional spatial flexibility to extend coverage, mitigate interference, and support aerial users or targets. These observations suggest a practical potential for partially upgrading existing wireless infrastructure with RAs rather than completely replacing conventional antenna arrays. However, the joint design and security of such hybrid antenna architectures in relay-assisted low-altitude networks remain insufficiently investigated~\cite{Zeng2026}.}

\subsection{Motivations and Contributions}
Motivated by the above, a decode‑and‑forward relay augmented with RA arrays is proposed. Both the base station and the relay are equipped with RA arrays and fixed‑position antenna arrays to simultaneously serve aerial and ground users. By partially enhancing conventional relay networks, the aim is to reduce deployment and hardware costs. Considering the security threat of unauthorized targets potentially intercepting confidential information in low‑altitude communications, the objective is to maximize the system secrecy rate (SR) through resource allocation under user quality‑of‑service constraints. The main contributions are summarized as follows:

\begin{itemize}
	
	\item A relay network with hybrid fixed‑position and rotatable antenna arrays is investigated. The deployment of rotatable arrays at conventional base stations and relay stations is considered. A SR maximization problem is formulated under constraints on transmit power, maximum zenith angle of the rotatable antennas, user quality of service (QoS), and path selection, with the aim of maximizing the network security performance through resource allocation, beamforming, and rotatable antenna orientation optimization.
	
	\item We provide optimization solutions for both \textcolor{black}{single ground–aerial user pair} and multi‑user cases. For the \textcolor{black}{single ground–aerial user pair} scenario, a leakage‑based sequential scheme is proposed. For the multi‑user case, a learning‑based approach is adopted to optimize the relay and BS antenna orientations and path selection, integrated with leakage theory, to maximize the SR.
	
	\item Simulation results validate the effectiveness of the proposed schemes. In the multi‑user scenario, a single base station alone cannot effectively cope with the multi‑user interference challenge. The additional alternative communication paths provided by the RIS can further reduce interference, thereby achieving higher SR performance. Moreover, the achievable SR of the proposed multi‑user scheme outperforms that of the conventional relay network equipped with isotropic antennas, and a twofold gain can be obtained.
\end{itemize}	
\subsection{Organization and Notation}

The remainder of the paper is organized as follows. Sec. \ref{sec:2} presents the system model. Secs. \ref{sec:3} and \ref{sec:4} investigate the \textcolor{black}{single ground–aerial user pair} and multi‑user solutions, respectively. Sec. \ref{sec:5} provides the simulation results, followed by the conclusions in Sec. \ref{sec:6}.

Notations: a, $\mathbf{a}$, and $\mathbf{A}$ denote a scalar, a vector, and a matrix, respectively. $\mathcal{A}$ denotes the set of elements. $\mathbb{C}^{a\times b}$ represents the set of $a\times b$ dimensional complex matrices. $(\cdot)^T$, $(\cdot)^H$, and $(\cdot)^{-1}$ denote the transpose, conjugate transpose, and matrix inverse operations, respectively. $|\cdot|$, $\|\cdot\|_2$, and $\|\cdot\|_{\infty}$ represent the absolute value, the $l_2$-norm, and the infinity norm, respectively. $\min\{\cdot\}$ and $\max\{\cdot\}$ denote the minimum and maximum operators, respectively. $[\cdot]_0^+$ denotes the positive part operator. $\Re(\cdot)$ and $\Im(\cdot)$ represent the real‑part and imaginary‑part extraction operations, respectively. \textcolor{black}{$\mathbf{a}[m]$ denotes the $m$-th element of $\mathbf{a}$.} $\mathbf{A}[a,:]$ and $\mathbf{A}[:,a]$ denote the extraction of the $a$-th row and the $a$-th column of a matrix, respectively. The superscript $^*$ denotes an optimized variable.
	\section{System Model and Problem Formulation}\label{sec:2}
\begin{figure}
	\centering
	\includegraphics[width=0.45\textwidth, trim = 10 30 2 2,clip]{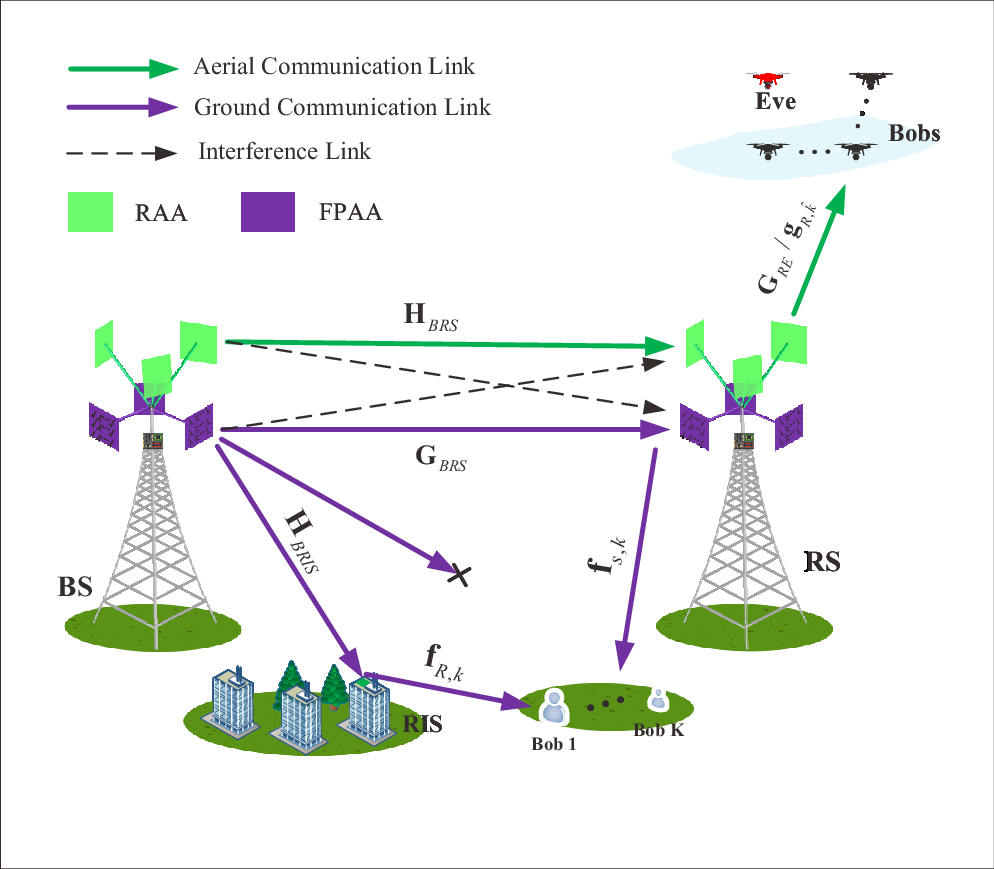}\\
	\caption{Illustration of a relay wireless network enabled by hybrid fixed-position and rotatable antenna arrays.}\label{fig:1}
\end{figure}
As illustrated in Fig. \ref{fig:1}, we consider a downlink communication scenario, which consists of a base station (BS), a decode‑and‑forward relay station (RS), an active RIS, an eavesdropper (Eve), $K$ ground legitimate users (Bobs), and $K_a$ aerial Bobs. The transmission is completed over two time slots with the aid of a decode-and-forward RS. In the first slot, a rotatable antenna array (RAA) and a fixed-position antenna array (FPAA) at the BS transmit signals to a RAA and a FPAA at the RS, respectively. In the second time slot, a RAA and a FPAA at the RS serve aerial and ground Bobs, respectively, and the BS can communicate with the ground Bobs via the RIS. Both the BS and the RS are equipped with $M_f$ FPAAs and $M_r$ RAAs. Each array is assumed to be equipped with $M$ antennas. The antenna spacing is set to uniform half-wavelength. 
The number of RIS elements is $Q$. Due to obstacle blockage and severe path loss, the BS transmits confidential information to the $K$ single-antenna ground Bobs via two primary paths: the first is the cascaded path reflected by the RIS, and the second is the path forwarded by the RS. Assuming that the distance between the RIS and the aerial Bobs is sufficiently large, the interference of RIS reflection can be neglected. \textcolor{black}{The $K_a$ single-antenna aerial Bobs receive their confidential information through the RS, while the direct links from the BS to the aerial Bobs are assumed to be negligible.} Owing to the broadcast nature of wireless networks, an Eve equipped with $E$ antennas is assumed to eavesdrop on the useful information corresponding to aerial Bobs. We also assume that Eve cannot intercept the confidential signals transmitted to ground Bobs. \textcolor{black}{It should be noted that the RIS is not intended to replace the relay. Rather, the relay remains the primary forwarding component, while the RIS serves as a complementary low‑cost communication path for ground users when the relay forwarding link suffers from severe interference or unfavorable propagation conditions.}

The received signals at the RS corresponding to the aerial and ground Bobs can be expressed as
\begin{align}&\mathbf{y}_{R,r}=\notag\\&\hat{\mathbf{U}}^H(\sqrt{ \frac{\rho P_0}{K_a}}\mathbf{H}_{BRS}\hat{\mathbf{W}}\hat{\mathbf{s}}+\sqrt{\tfrac{(1-\rho)P_0}{K-\tilde{K}}}\hat{\mathbf{G}}_{BRS}{\mathbf{W}}\mathbf{O}{\mathbf{s}}+\mathbf{n}_1),\\
&\mathbf{y}_{F,r}=\notag\\&{\mathbf{U}}^H(\sqrt{\tfrac{(1-\rho)P_0}{K-\tilde{K}}}{\mathbf{G}}_{BRS}{\mathbf{W}}\mathbf{O}{\mathbf{s}}+\sqrt{ \frac{\rho P_0}{K_a}}\hat{\mathbf{H}}_{BRS}\hat{\mathbf{W}}\hat{\mathbf{s}}+\mathbf{n}_2),\end{align}
respectively. $\tilde K$ denotes the number of Bobs communicating via the RIS reflection path. 
$P_0$ denotes the transmit power. 
\textcolor{black}{$0\leq\rho\leq1$} represents the power allocation factor. $\mathbf{H}_{BRS}\in\mathbb{C}^{M\times M}$ and $\hat{\mathbf{G}}_{BRS}\in\mathbb{C}^{M\times M}$ represent the desired and interference channels from BS to RS for the aerial Bobs, respectively. $\mathbf{G}_{BRS}\in\mathbb{C}^{M\times M}$ and $\hat{\mathbf{H}}_{BRS}\in\mathbb{C}^{M\times M}$ denote the desired and interference channels from BS to RS for the ground Bobs, respectively. $\hat{\mathbf{s}}=[\hat{s}_1,\hat{s}_2,\ldots,\hat{s}_{K_a}]^T\in\mathbb{C}^{K_a\times1}\ (\hat{k}\in\mathcal{K}_1=\{1,2,\ldots,K_a\})$ and $\mathbf{s}=[s_1,s_2,\ldots,s_{{K}}]^T\in\mathbb{C}^{{K}\times1}\ (k\in\mathcal{K}_2=\{1,2,\ldots,{K}\})$ denote the signals transmitted to the aerial and ground Bobs, respectively, each with a statistical expectation of one. ${\mathbf{U}}=[\mathbf{u}_1,\mathbf{u}_2,\ldots,\mathbf{u}_{{K}}]\in\mathbb{C}^{M\times {K}}$ and $\hat{\mathbf{U}}=[\hat{\mathbf{u}}_1,\hat{\mathbf{u}}_2,\ldots,\hat{\mathbf{u}}_{K_a}]\in\mathbb{C}^{M\times K_a}$ denote the receive beamforming matrices. We consider $\mathbf{U}^H\mathbf{U}=\mathbf{I}_{{K}}$ and $\hat{\mathbf{U}}^H\hat{\mathbf{U}}=\mathbf{I}_{K_a}$. $\mathbf{n}_1\sim\mathcal{CN}(0,\sigma_{1}^{2}\mathbf{I}_{M})$ is the noise experienced by RAAs. $\mathbf{n}_2\sim\mathcal{CN}(0,\sigma_{2}^{2}\mathbf{I}_{M})$ denotes the noise experienced by FPAAs. $\mathbf{W}=[\mathbf{w}_1,\mathbf{w}_2,\ldots,\mathbf{w}_{K}]\in\mathbb{C}^{M\times {K}}$ and $\hat{\mathbf{W}}=[\hat{\mathbf{w}}_1,\hat{\mathbf{w}}_2,\ldots,\hat{\mathbf{w}}_{K_a}]\in\mathbb{C}^{M\times K_a}$ denote the transmit beamforming matrices for the ground and aerial Bobs, respectively. $\mathbf{O}=\mathrm{diag}(\zeta_1,\zeta_2,\ldots,\zeta_K)$, $\zeta_k\in\{0,1\}$ is a binary variable used for transmission path selection. \textcolor{black}{Note that the mathematical models for the cases where all ground users are served by the RIS and where no aerial users exist can be readily derived, respectively. Therefore, division‑by‑zero errors are not a concern.}

The signal ${{y}}_k$ received by the $k$-th ground Bob can be expressed as
\begin{align}{{y}}_k&=\sqrt{{\mu}_k P_s}\zeta_k\mathbf{f}_{s,k}^H{\mathbf{u}}_{t,k}{s}_k\!+\!\sqrt{\frac{P_0}{\tilde K}}(1\!\!-\!\!\zeta_k)\mathbf{f}_{R,k}^H\mathbf{\Theta}\mathbf{H}_{BRIS}{\mathbf{v}}_{t,k}{s}_k\notag\\&+\sum_{i=1,i\neq k}^{K}\Big(\sqrt{{\mu}_i P_s}\zeta_i\mathbf{f}_{s,k}^H{\mathbf{u}}_{t,i}{s}_i\notag\\&\qquad\qquad\quad+\sqrt{\frac{P_0}{\tilde K}}(1-\zeta_i)\mathbf{f}_{R,k}^H\mathbf{\Theta}\mathbf{H}_{BRIS}{\mathbf{v}}_{t,i}{s}_i\Big)\notag\\&+(1-\zeta_k)\mathbf{f}_{R,k}^H\mathbf{\Theta}\mathbf{n}_r+{n}_k,\end{align}
where $P_s$ denotes the transmit power corresponding to RS. 
 $\mathbf{v}_{t,k}\in\mathbb{C}^{M\times 1}$ denotes the transmit beamforming vector. ${\mu}_k$ denotes the power allocation factor. $\mathbf{f}_{R,k}\in\mathbb{C}^{Q\times 1}$ and $\mathbf{f}_{s,k}\in\mathbb{C}^{M\times 1}$ denote the desired channels from the RIS and the RS to the $k$-th ground Bob, respectively. $\mathbf{H}_{BRIS}\in\mathbb{C}^{Q\times M}$ represents the desired channel from the BS to the RIS. ${\mathbf{u}}_{t,k}\in\mathbb{C}^{M\times 1}$ denotes the transmit beamforming vector at the RS for the $k$-th ground Bob. ${n}_k\sim\mathcal{CN}(0,\sigma_{3}^{2})$ and $\mathbf{n}_r\sim\mathcal{CN}(0,\sigma_{4}^{2}\mathbf{I}_{Q})$ are the noise. $\mathbf{\Theta}=\mathrm{diag}([\alpha_1 e^{j\beta_1},\alpha_2 e^{j\beta_2},\ldots,\alpha_Q e^{j\beta_Q}]^T)$ is the phase shift matrix of the RIS, where $\alpha_q\in(1,\alpha_{\mathrm{max}}]$ and $\beta_q\in(0,2\pi]$ $(q\in\mathcal{Q}=\{1,2,\ldots,Q\})$ are the amplitude and phase of the $q$-th RIS element, respectively. $\alpha_{\mathrm{max}}$ is the maximum reflection amplitude.

Correspondingly, the signal $\hat{y}_{\hat{k}}$ received by the $\hat{k}$-th aerial Bob can be expressed as
\begin{equation}\hat{{y}}_{\hat{k}}=\sqrt{\hat{\mu}_{\hat{k}} P_s}\mathbf{g}_{r,\hat{k}}^H\hat{\mathbf{u}}_{t,\hat{k}}\hat{s}_{\hat{k}}+\sum_{i=1,i\neq \hat{k}}^{K_a}\sqrt{\hat{\mu}_{i} P_s}\mathbf{g}_{r,\hat{k}}^H\hat{\mathbf{u}}_{t,i}\hat{s}_{i}+\hat{n}_{\hat{k}},\end{equation}
where $\hat{\mu}_{\hat{k}}$ denotes the power allocation factor for the $\hat{k}$-th aerial Bob. $\mathbf{g}_{r,\hat{k}}\in\mathbb{C}^{M\times 1}$ and $\hat{\mathbf{u}}_{t,\hat{k}}\in\mathbb{C}^{M\times 1}$ denote the desired channel and transmit beamforming vector from the RS to the $\hat{k}$-th aerial Bob, respectively. $\hat{n}_{\hat{k}}\sim\mathcal{CN}(0,\sigma_{5}^{2})$ is the noise. The transmit power of the RS satisfies \begin{equation}\label{4}\sum_{\hat{k}=1}^{K_a}\hat{\mu}_{\hat{k}}+\sum_{k=1}^{{K}}\zeta_k{\mu}_k=1.\end{equation}

For Eve, the received signal ${y}_{E,\hat{k}}$ can be expressed as
\begin{align}{y}_{E,\hat{k}}&=\sqrt{\hat{\mu}_{\hat{k}} P_s}\tilde{\mathbf{u}}_{\hat{k}}^H\mathbf{G}_{RE}^H\hat{\mathbf{u}}_{t,\hat{k}}\hat{s}_{\hat{k}}\notag\\&+\sum_{i=1,i\neq\hat{k}}^{K_a}\sqrt{\hat{\mu}_{i} P_s}\tilde{\mathbf{u}}_{\hat{k}}^H\mathbf{G}_{RE}^H\hat{\mathbf{u}}_{t,i}\hat{s}_{i}+\tilde{\mathbf{u}}_{\hat{k}}^H\tilde{\mathbf{n}},\end{align}
where $\mathbf{G}_{RE}\in\mathbb{C}^{M\times E}$ and $\tilde{\mathbf{n}}\sim\mathcal{CN}(0,\sigma_{6}^{2}\mathbf{I}_E)$ denote the desired channel and the noise from the RS to Eve, respectively. $\tilde{\mathbf{u}}_{\hat{k}}\in\mathbb{C}^{E\times 1}$ denote the receive beamforming matrix. Assuming that the number of antennas $E< M$ equipped at the Eve is larger than the number of Bobs $K_a$, enabling it to demodulate up to $K_a$ independent data streams. 

\subsection{Channel Model}
For ground and aerial Bobs operating in distinct propagation environments, typical channel models are considered. Ground Bobs, subject to blockages such as trees and buildings, can be modeled as multipath channels. For aerial Bobs, the line-of-sight (LoS) channel is dominant.
The omnidirectional antennas of the FPAAs are fixed in position. The directional gain pattern for the $m_2$-th ($m_2\in\mathcal{M}=\{1,2,\ldots,M\}$) RA corresponding to the BS can be denoted as
\begin{align}&B_{m_2,\hat{m}}^{BS}=\notag\\&\begin{cases}2(2p+1)\cos^{2p}(\epsilon_{m_2,\hat{m}}^{BS}),&\epsilon_{m_2,\hat{m}}^{BS}\in\left[0,\frac{\pi}{2}\right),\varphi_{m_2,\hat{m}}^{BS}\in[0,2\pi)\\0,&\text{otherwise,}\end{cases}\notag\\&\triangleq\begin{cases}B_0[(\vec{\mathbf{p}}_{m_2}^{BS})^T\vec{\mathbf{d}}_{m_2}^{\hat{m}}]_+^{2p},&\theta_{m_2}^{BS}\in\left[0,\theta_{\max}\right),\vartheta_{m_2}^{BS}\in[0,2\pi)\\0,&\text{otherwise,}\end{cases}\label{6}\end{align}
where $(\epsilon_{m_2}^{BS},\varphi_{m_2}^{BS})$ denotes a pair of incident angles, $B_{m_2,\hat{m}}^{BS}\triangleq B_{m_2,\hat{m}}^{BS}(\epsilon_{m_2,\hat{m}}^{BS},\varphi_{m_2,\hat{m}}^{BS})$. $\theta_{m_2}^{BS}$ and $\vartheta_{m_2}^{BS}$ denote the zenith angle and azimuth angle, respectively, in the local coordinate system of the $m_2$-th RA. $\theta_{\max}$ represents the maximum zenith angle. $p\geq 0$ denotes the directivity factor. $\vec{\mathbf{p}}_{m_2}^{BS}$ represents the pointing vector of the $m_2$-th RA at the BS, given by
\begin{align}\label{7}
	\vec{\mathbf{p}}_{m_2}^{BS}=[\sin\theta_{m_2}^{BS}\cos\vartheta_{m_2}^{BS},\sin\theta_{m_2}^{BS}\sin\vartheta_{m_2}^{BS},\cos\theta_{m_2}^{BS}]^T.
\end{align}
 $\vec{\mathbf{d}}_{m_2}^{\hat{m}}=[x_{m_2}^{\hat{m}},y_{m_2}^{\hat{m}},z_{m_2}^{\hat{m}}]^T$ denotes the pointing vector from the $m_2$-th RA at the BS to the 
$\hat{m}$-th $(\hat{m}\in\{\bar{m}_1,\bar{m}_2\})$ antenna at the RS, $\bar{m}_1\in\mathcal{M}$ and $\bar{m}_2\in\mathcal{M}$. Similarly, $B_{\tilde{m},\bar{m}_2}^{RS,1}\triangleq B_{\tilde{m},\bar{m}_2}^{RS,1}(\epsilon_{\tilde{m},\bar{m}_2}^{RS,1},\varphi_{\tilde{m},\bar{m}_2}^{RS,1})$, $B_{\bar{m}_2,\hat{k}}^{RS,2}\triangleq B_{\bar{m}_2,\hat{k}}^{RS,2}(\epsilon_{\bar{m}_2,\hat{k}}^{RS,2},\varphi_{\bar{m}_2,\hat{k}}^{RS,2})$, and $B_{\bar{m}_2,e}^{RS,2}\triangleq B_{\bar{m}_2,e}^{RS,2}(\epsilon_{\bar{m}_2,e}^{RS,2},\varphi_{\bar{m}_2,e}^{RS,2})$ denote the directional gain pattern for the $\bar{m}_2$-th RA corresponding to the RS,  where $\tilde{m}\in \{m_1,m_2\}$. $\vec{\mathbf{k}}_{\bar{m}_2}^{\hat{k}}=[x_{\bar{m}_2}^{\hat{k}},y_{\bar{m}_2}^{\hat{k}},z_{\bar{m}_2}^{\hat{k}}]^T$ denotes the pointing vector from the $\bar{m}_2$-th RA at the RS to the $\hat{k}$-th aerial Bob. In the first time slot, the pointing vector of the $\bar{m}_2$-th RA at the RS can be given by
\begin{align}\label{8}
	\vec{\mathbf{p}}_{\bar{m}_2}^{RS,1}=[\sin\theta_{\bar{m}_2}^{RS,1}\cos\vartheta_{\bar{m}_2}^{RS,1},\sin\theta_{\bar{m}_2}^{RS,1}\sin\vartheta_{\bar{m}_2}^{RS,1},\cos\theta_{\bar{m}_2}^{RS,1}]^T.
\end{align}By setting $p=0$, we obtain the gain pattern of the FPA. Accordingly, $	\vec{\mathbf{p}}_{\bar{m}_2}^{RS,2}$ denotes the pointing vector corresponding to the $\bar{m}_2$-th antenna at the RS in the second time slot. $\theta_{\bar{m}_2}^{RS,1}$ and $\vartheta_{\bar{m}_2}^{RS,1}$ denote the zenith angle and azimuth angle at the RS in the first time slot, respectively. $\theta_{\bar{m}_2}^{RS,2}$ and $\vartheta_{\bar{m}_2}^{RS,2}$ denote the zenith angle and azimuth angle at the RS in the second time slot, respectively.

The channel power gains between the $m_1$-th ($m_1\in\mathcal{M}$) FPA at the BS and the $\bar{m}_1$-th FPA at the RS, between the $m_1$-th FPA at the BS and the 
$\bar{m}_2$-th RA at the RS, between the $m_2$-th RA at the BS and the $\bar{m}_1$-th FPA at the RS, and between the $m_2$-th RA at the BS and the $\bar{m}_2$-th RA at the RS can be expressed as
\begin{align}
	&A_{m_1,\bar{m}_1}^{FF}=\frac{D}{\pi(r_{m_1,\bar{m}_1}^{FF})^2},\\
	&A_{m_1,\bar{m}_2}^{FR}(\vec{\mathbf{p}}_{\bar{m}_2}^{RS,1})=\frac{D}{2\pi(r_{m_1,\bar{m}_2}^{FR})^2}B_{m_1,\bar{m}_2}^{RS,1},\\
	&A_{m_2,\bar{m}_1}^{RF}(\vec{\mathbf{p}}_{m_2}^{BS})=\frac{D}{2\pi(r_{m_2,\bar{m}_1}^{RF})^2}B_{m_2,\bar{m}_1}^{BS},\\
	&A_{m_2,\bar{m}_2}^{RR}(\vec{\mathbf{p}}_{m_2}^{BS},\vec{\mathbf{p}}_{\bar{m}_2}^{RS,1})=\frac{D}{4\pi(r_{m_2,\bar{m}_2}^{RR})^2}B_{m_2,\bar{m}_2}^{BS}B_{m_2,\bar{m}_2}^{RS,1},
\end{align}
respectively. $D$ denotes the antenna size. $r_{m_1,\bar{m}_1}^{FF}$, $r_{m_1,\bar{m}_2}^{FR}$, $r_{m_2,\bar{m}_1}^{RF}$, and $r_{m_2,\bar{m}_2}^{RR}$ denote the distances between the $m_1$-th FPA at the BS and the $\bar{m}_1$-th FPA at the RS, between the $m_1$-th FPA at the BS and the 
$\bar{m}_2$-th RA at the RS, between the $m_2$-th RA at the BS and the $\bar{m}_1$-th FPA at the RS, and between the $m_2$-th RA at the BS and the $\bar{m}_2$-th RA at the RS. Similarly, the channel power gains from the RS to aerial Bob, ground Bob, and Eve are denoted as $A_{\bar{m}_2,{\hat{k}}}^{RB}(\vec{\mathbf{p}}_{\bar{m}_2}^{RS})$, $A_{\bar{m}_1,{{k}}}^{FB}$, and $A_{\bar{m}_2,e}^{RE}(\vec{\mathbf{p}}_{\bar{m}_2}^{RS})$, respectively. $e\in\mathcal{E}=\{1,2,\ldots,E\}$ denotes the index of the Eve's antenna. The channel power gain from the BS to RIS can be represented as $A_{m_1,q}^{Fq}$.

Then, we have
\begin{align}
	&\!\!\mathbf{H}_{BRS}[\bar{m}_2,m_2] = \sqrt{A_{m_2,\bar{m}_2}^{RR}(\vec{\mathbf{p}}_{m_2}^{BS},\vec{\mathbf{p}}_{\bar{m}_2}^{RS,1})}e^{-j\frac{2\pi}{\lambda}(r_{m_2,\bar{m}_2}^{RR}-r_{B})},\end{align} \begin{align}
	&\!\!\hat{\mathbf{H}}_{BRS}[\bar{m}_1,m_2] = \sqrt{A_{m_2,\bar{m}_1}^{RF}(\vec{\mathbf{p}}_{m_2}^{BS})}e^{-j\frac{2\pi}{\lambda}(r_{m_2,\bar{m}_1}^{RF}-r_{B})},\\
	&\!\!\mathbf{G}_{BRS}[\bar{m}_1,m_1] = \sqrt{A_{m_1,\bar{m}_1}^{FF}}e^{-j\frac{2\pi}{\lambda}(r_{m_1,\bar{m}_1}^{FF}-r_{B})},\\	&\!\!\hat{\mathbf{G}}_{BRS}[\bar{m}_2,m_1] = \sqrt{A_{m_1,\bar{m}_2}^{FR}(\vec{\mathbf{p}}_{\bar{m}_2}^{RS,1})}e^{-j\frac{2\pi}{\lambda}(r_{m_1,\bar{m}_2}^{FR}-r_{B})},	\\
	&\!\!\mathbf{f}_{s,k}[\bar{m}_1] = \sqrt{A_{\bar{m}_1,{{k}}}^{FB}}e^{-j\frac{2\pi}{\lambda}(r_{\bar{m}_1,k}^{RB}-r_{k})},\\
	&\!\!\mathbf{g}_{r,\hat{k}}[\bar{m}_2] = \sqrt{A_{\bar{m}_2,{\hat{k}}}^{RB}(\vec{\mathbf{p}}_{\bar{m}_2}^{RS,2})}e^{-j\frac{2\pi}{\lambda}(r_{\bar{m}_2,\hat{k}}^{RB}-r_{\hat{k}})},\\
	&\!\!\mathbf{G}_{RE}[\bar{m}_2,e] = \sqrt{A_{\bar{m}_2,e}^{RE}(\vec{\mathbf{p}}_{\bar{m}_2}^{RS,2})}e^{-j\frac{2\pi}{\lambda}(r_{\bar{m}_2,e}^{RE}-r_{E})},
\end{align}
where $r_{\bar{m}_1,k}^{RB}$, $r_{\bar{m}_2,\hat{k}}^{RB}$, and $r_{\bar{m}_2,e}^{RE}$ denote the distances from the RS to the $k$-th ground Bob, the $\hat{k}$ aerial Bob, and Eve. $r_{B}$, $r_{k}$, $r_{\hat{k}}$, and $r_{E}$ denote the reference distances between the RS and the BS, between the $k$-th ground Bob and the RS, between the $\hat{k}$-th aerial Bob and the RS, and between Eve and the RS, respectively. $\mathbf{H}_{BRIS}$ and $\mathbf{f}_{R,k}$ consist of LoS and non-line-of-sight (NLoS) components and can be expressed as~\cite{Zheng2026}
\begin{align}
	&\mathbf{H}_{BRIS}[q,m_1] = \sqrt{A_{m_1,q}^{Fq}}e^{-j\frac{2\pi}{\lambda}(r_{m_1,q}^{Fq}-r_{B})}\notag\\
	&\qquad\qquad+\sum_{l=1}^{L}\frac{\sqrt{\sigma_{l}A_{m_1,l}^{FS}}}{r_{l,q}^{SR}}e^{-j\frac{2\pi}{\lambda}(r_{m_1,l}^{SR}+r_{l,q}^{SR}-r_{B})+j\chi_{l}},\\
	&\mathbf{f}_{R,k}[q] = \sqrt{A_{q,k}^{qB}}e^{-j\frac{2\pi}{\lambda}(r_{q,k}^{qB}-r_{C})}\notag\\
	&\qquad\qquad+\sum_{\bar{l}=1}^{\bar{L}}\frac{\sqrt{\bar{\sigma}_{\bar{l}}A_{q,\bar{l}}^{qS}}}{r_{\bar{l},k}^{SB}}e^{-j\frac{2\pi}{\lambda}(r_{q,\bar{l}}^{SB}+r_{\bar{l},k}^{SB}-r_{C})+j\bar{\chi}_{\bar{l}}}.
\end{align}
respectively. Here, $\sigma_{l}$ and $\bar{\sigma}_{\bar{l}}$ are the radar cross sections. $r_{m_1,q}^{Fq}$, $r_{m_1,l}^{SR}$, $r_{l,q}^{SR}$, $r_{q,k}^{qB}$, $r_{q,\bar{l}}^{SB}$, and $r_{\bar{l},k}^{SB}$ represent the distances between the $m_1$-th FPA at BS and the $q$-th RIS element,  between the $m_1$-th FPA at BS and the $l$-th scatterer, between the $l$-th scatterer and the $q$-th RIS element, between the $q$-th RIS element and the $k$-th Bob, between the $q$-th RIS element and the $\bar{l}$-th scatterer, and between the $\bar{l}$-th scatterer and the $k$-th Bob. $A_{m_1,q}^{Fq}$, $A_{m_1,l}^{FS}$, $A_{q,k}^{qB}$, and $A_{q,\bar{l}}^{qS}$ are the channel power gains of the corresponding paths. $r_C$ denotes the reference distance between RIS and Bob. $\chi_{{l}}$ and $\bar{\chi}_{\bar{l}}$ represent the phase shifts introduced by the scatterer clusters.

Note that the far field is an approximation of the near field, and the presented channel models can be readily extended to far‑field scenarios. Near‑field channels are considered due to the large virtual aperture of the distributed transceiver arrays and the deployment of the RIS in close proximity to ground Bobs~\cite{Zhu2025, Xiao2024c}.
\subsection{Performance Metrics}
The SR is considered as the performance metric for evaluating the network. The SINR corresponding to RS can be represented as 
\begin{align}
	\gamma_r = \min\{\gamma_{r,1},\gamma_{r,2}\},
\end{align}
where
\begin{align}
&\gamma_{r,1}=\frac{\frac{\rho P_0}{K_a}\|\hat{\mathbf{U}}^H\mathbf{H}_{BRS}\hat{\mathbf{W}}\hat{\mathbf{s}}\|^2}{\frac{(1-\rho) P_0}{K-\tilde K}\|\hat{\mathbf{U}}^H\hat{\mathbf{G}}_{BRS}{\mathbf{W}}\mathbf{O}{\mathbf{s}}\|^2+\sigma_1^2},
\end{align} \begin{align}&\gamma_{r,2}=\frac{\frac{(1-\rho) P_0}{K-\tilde K}\|\mathbf{U}^H\mathbf{G}_{BRS}{\mathbf{W}}\mathbf{O}\mathbf{s}\|^2}{\frac{\rho P_0}{K_a}\|\mathbf{U}^H\hat{\mathbf{H}}_{BRS}\hat{\mathbf{W}}\hat{\mathbf{s}}\|^2+\sigma_2^2}.
\end{align}
The SINR of the $k$-th ground Bob can be expressed as 
\begin{align}
	&\gamma_k = \frac{{\mu}_k P_s\zeta_k|\mathbf{f}_{s,k}^H{\mathbf{u}}_{t,k}|^2+\frac{P_0}{\tilde K}(1-\zeta_k)|\mathbf{f}_{R,k}^H\mathbf{\Theta}\mathbf{H}_{BRIS}{\mathbf{v}}_{t,k}|^2}{N_0+\sigma_3^2+(1-\zeta_k)\sigma_4^2\|\mathbf{f}_{R,k}^H\mathbf{\Theta}\|^2},
\end{align}
where $N_0 = \sum_{i=1,i\neq k}^{K}(\frac{P_0}{\tilde K}(1-\zeta_i)|\mathbf{f}_{R,k}^H\mathbf{\Theta}\mathbf{H}_{BRIS}{\mathbf{v}}_{t,i}|^2+{\mu}_i P_s\zeta_i|\mathbf{f}_{s,k}^H{\mathbf{u}}_{t,i}|^2)$. Then, the achievable rate of the $k$-th ground Bob can be denoted as
\begin{align}
	&R_k = \min\{\log_2(1+\gamma_k),\log_2(1+\gamma_{r,2})\}.
\end{align}
The SINR of the $\hat{k}$-th aerial Bob can be denoted as 
\begin{align}
	&\hat{\gamma}_{\hat{k}} = \frac{\hat{\mu}_{\hat{k}} P_s|\mathbf{g}_{r,\hat{k}}^H\hat{\mathbf{u}}_{t,\hat{k}}|^2}{\sum_{i=1,i\neq \hat{k}}^{K_a}{\hat{\mu}_{i} P_s}|\mathbf{g}_{r,\hat{k}}^H\hat{\mathbf{u}}_{t,i}|^2+\sigma_5^2}.
\end{align}
Then, the achievable rate of the $k$-th \textcolor{black}{aerial} Bob can be denoted as
\begin{align}
	&\hat{R}_{\hat k} = \min\{\log_2(1+\hat{\gamma}_{\hat k}),\log_2(1+\gamma_{r,1})\}.
\end{align}
The SINR of Eve can be expressed as 
\begin{align}
	&\hat{\gamma}_{Eve,\hat{k}} = \frac{\hat{\mu}_{\hat{k}} P_s|\tilde{\mathbf{u}}_{\hat{k}}^H\mathbf{G}_{RE}^H\hat{\mathbf{u}}_{t,\hat{k}}|^2}{\sum_{i=1,i\neq\hat{k}}^{K_a}{\hat{\mu}_{i} P_s}|\tilde{\mathbf{u}}_{\hat{k}}^H\mathbf{G}_{RE}^H\hat{\mathbf{u}}_{t,i}|^2+\sigma_6^2}.
\end{align}
For all signals that Eve may demodulate, the maximum SINR can be denoted as $\hat{\gamma}_{\max}=\|[\hat{\gamma}_{Eve,1},\hat{\gamma}_{Eve,2},\ldots,\hat{\gamma}_{Eve,K_a}]^T\|_{\infty}$. The worst‑case security performance is considered. Then, the achievable SR can be expressed as
\begin{align}
	R_s = \max\{{\hat{R}_{\hat k}^{\min} }-\log_2(1+\hat{\gamma}_{\max}),0\},
\end{align}
where $\hat{R}_{\hat k}^{\min}$ denotes the minimum rate corresponding to the aerial Bobs.
\subsection{Problem Formulation}
The objective is to maximize the achievable SR. The network has a limited power budget, satisfying $\eqref{4}$. To maximize the SR, all transmit power is used for signal transmission. To guarantee the QoS for Bobs, the transmission rate from the BS to the RS is no less than that from the RS to the Bobs, as given by $\gamma_{r,1}\geq\hat{\gamma}_{\hat{k}}$ and $\gamma_{r,2}\geq\gamma_k$. The adjustable range of the RAs is limited, as shown by $\theta_{m_2}^{BS}\in[0,\theta_{\operatorname*{max}}]$, $\theta_{\bar{m}_2}^{RS,1}\in[0,\theta_{\operatorname*{max}}]$, and $\theta_{\bar{m}_2}^{RS,2}\in[0,\theta_{\operatorname*{max}}]$. The adjustable amplitude of the active RIS is limited, i.e., $\alpha_q\in(1,\alpha_{\mathrm{max}}]$. For transmission path selection, we have $\zeta_k\in\{0,1\}$. For clarity, define the set of variables \begin{align}&\mathbb{A}=\{\textcolor{black}{\rho},{\mu}_1,\ldots,{\mu}_K,\hat{\mu}_1,\ldots,\hat{\mu}_{K_a}\},\\
	&\mathbb{B}=\{{\zeta}_1,\ldots,{\zeta}_K\},\\
	&\mathbb{P}=\{\vec{\mathbf{p}}_{1}^{BS},\ldots,\vec{\mathbf{p}}_{M}^{BS},\vec{\mathbf{p}}_{1}^{RS,1},\ldots,\vec{\mathbf{p}}_{M}^{RS,1},\vec{\mathbf{p}}_{1}^{RS,2},\ldots,\vec{\mathbf{p}}_{M}^{RS,2}\},\\
	&\mathbb{U}=\{\mathbf{U},\hat{\mathbf{U}},\tilde{\mathbf{u}}_{1},\ldots,\tilde{\mathbf{u}}_{K_a}\},\\
	&\mathbb{V}=\{\mathbf{u}_{t,1},\ldots,\mathbf{u}_{t,K},\hat{\mathbf{u}}_{t,1},\ldots,\hat{\mathbf{u}}_{t,K_a},\mathbf{v}_{t,1},\ldots,\mathbf{v}_{t,K}\},\\
	&\mathbb{W}=\{\mathbf{W},\hat{\mathbf{W}}\}.
\end{align}
Then, the optimization problem can be formulated as
 \begin{subequations}
\begin{align}\mathrm{P1}:&{\operatorname*{max}_{\mathbb{A},\mathbb{B},\mathbb{P},\mathbb{U},\mathbb{V},\mathbb{W},\bm{\Theta}}}\quad R_s\label{s}\\\mathrm{s.t.}\ &\begin{cases}\color{black}0\leq\rho\leq1,\\ \sum_{\hat{k}=1}^{K_a}\hat{\mu}_{\hat{k}}+\sum_{k=1}^{{K}}\zeta_k{\mu}_k=1, \color{black}{\hat{\mu}_{\hat{k}}\geq0, {\mu}_k\geq0, \forall \hat{k}, k,}\end{cases}\label{s1}\\
	&\gamma_{r,1}\geq\hat{\gamma}_{\hat{k}}\geq\nu,\gamma_{r,2}\geq\gamma_k\geq\nu,\forall \hat{k},k,\label{s2}\\
	&\begin{cases}0\leq\theta_{m_2}^{BS}\leq\theta_{\operatorname*{max}},\forall m_2,\\0\leq\theta_{\bar{m}_2}^{RS,1}\leq\theta_{\operatorname*{max}},0\leq\theta_{\bar{m}_2}^{RS,2}\leq\theta_{\operatorname*{max}},\forall  \bar{m}_2,\end{cases}\label{s3}\\
	&\zeta_k\in\{0,1\},\forall k,\label{s4}\\
	&\alpha_q\in(1,\alpha_{\mathrm{max}}],\forall q,\label{s5}\\
	&\|\mathbf{w}_{k}\|=1,\|\hat{\mathbf{w}}_{\hat{k}}\|=1,\forall k,\forall \hat{k},\label{s6}\\
	&\|\mathbf{u}_{t,k}\|=1,\|\mathbf{v}_{t,k}\|=1,\|\hat{\mathbf{u}}_{t,\hat{k}}\|=1,\forall k,\forall \hat{k},\label{s7}\\
	&\mathbf{U}^H\mathbf{U}=\mathbf{I}_{{K}},\hat{\mathbf{U}}^H\hat{\mathbf{U}}=\mathbf{I}_{K_a},\|\tilde{\mathbf{u}}_{\hat{k}}\|=1,\forall \hat{k}.\label{s8}\end{align}
 \end{subequations}
 P1 is a non‑convex optimization problem with coupled multiple variables. \textcolor{black}{\eqref{s1} denotes the power allocation at the BS and the RS.} \eqref{s2} ensures the QoS requirements of the Bobs. $\nu$ denotes the minimum SINR threshold. \eqref{s3} denotes the adjustable range of the zenith angles of the RAs. $\mathbb{B}$ is a set of binary variables. \eqref{s6}–\eqref{s8} represent normalization operations. Consequently, P1 is generally difficult to solve directly.
\section{\textcolor{black}{Single Ground–Aerial User Pair Case with Stepwise Optimization}}\label{sec:3}
\textcolor{black}{In this section, a single ground–aerial user pair case is considered, i.e., $K=1$ and $K_a=1$, corresponding to one ground Bob and one aerial Bob, respectively. This simplified setting eliminates intra-group multi-user interference and facilitates the analysis of the fundamental coupling between the ground and aerial communication links.}
 When $\zeta_k=0$, a single Bob mainly receives signals via the RIS reflection path, which is a typical scenario for RIS‑enabled communication. When $\zeta_k=1$, the Bob receives signals forwarded by the hybrid fixed‑position and RA arrays at the RS. The case $\zeta_k=1$ is studied in detail. 
Specifically, under the QoS guarantee, \eqref{s2} can be denoted as
\begin{align}
f_1&=\frac{\rho P_0|\hat{\mathbf{u}}_{\hat{k}}^H\mathbf{H}_{BRS}\hat{\mathbf{w}}_{\hat{k}}|^2}{(1-\rho) P_0|\hat{\mathbf{u}}_{\hat{k}}^H{\hat{\mathbf{G}}}_{BRS}{\mathbf{w}}_{{k}}|^2+\sigma_1^2}\notag\\&\geq\frac{\hat{\mu}_{\hat{k}} P_s}{\sigma_5^2}|\mathbf{g}_{r,\hat{k}}^H\hat{\mathbf{u}}_{t,\hat{k}}|^2=\nu,\label{34}\\
f_2&=\frac{(1-\rho) P_0| \mathbf{u}_k^H\mathbf{G}_{BRS}{\mathbf{w}_k}|^2}{\rho P_0|\mathbf{u}_k^H\hat{\mathbf{H}}_{BRS}\hat{\mathbf{w}}_{\hat{k}}|^2+\sigma_2^2}
\geq \frac{{\mu}_k P_s}{\sigma_3^2}|\mathbf{f}_{s,k}^H{\mathbf{u}}_{t,k}|^2=\nu.\label{35}
\end{align}
Based on the principle of maximizing the received SINR, we consider that the initial directions of the RAs are oriented toward their respective targets~\cite{Zheng2026}. The RAs at the BS are directed toward the RAA at the RS. During the first time slot, the receiving RAs at the RS are oriented toward the RAA at the BS. During the second time slot, the RAs at the RS are oriented toward a reference Bob. Correspondingly, the initial values of the pointing vectors are $\vec{\mathbf{d}}_{m_2}^{1}=[x_{m_2}^{1},y_{m_2}^{1},z_{m_2}^{1}]^T$, $\vec{\mathbf{d}}_{1}^{\bar{m}_2}=[x^{\bar{m}_2}_{1},y^{\bar{m}_2}_{1},z^{\bar{m}_2}_{1}]^T$, and $\vec{\mathbf{k}}_{\bar{m}_2}^{1}=[x_{\bar{m}_2}^{1},y_{\bar{m}_2}^{1},z_{\bar{m}_2}^{1}]^T$, respectively. Then, substituting \eqref{34} into \eqref{s}, we have
 \begin{subequations}
	\begin{align}\mathrm{P2}:&{\operatorname*{max}_{\mathbb{A},\mathbb{B},\mathbb{P},\mathbb{U},\mathbb{V},\mathbb{W}}}\quad \frac{\sigma_6^2+\sigma_6^2\nu}{\sigma_6^2+(1-\mu_k)P_s|\tilde{\mathbf{u}}_{\hat{k}}^H\mathbf{G}_{RE}^H\hat{\mathbf{u}}_{t,\hat{k}}|^2}\label{36a}\\&\mathrm{s.t.}\ \eqref{s3},\eqref{s5},\eqref{s6},\eqref{s7},\eqref{s8},\eqref{34},\eqref{35},
	\end{align}
\end{subequations}
To maximize the ground Bob's transmission rate, according to the maximum ratio transmission (MRT) criterion, we have ${\mathbf{u}}_{t,k}^*=\frac{\mathbf{f}_{s,k}}{\|\mathbf{f}_{s,k}\|}$. 
$\hat{\mathbf{u}}_{t,\hat{k}}^*$ can be designed using null-space projection (NSP) to prevent confidential information leakage, i.e.,
\begin{align}
	\hat{\mathbf{u}}_{t,\hat{k}}^* =(\mathbf{I}_M- \mathbf{G}_{RE}(\mathbf{G}_{RE}^H\mathbf{G}_{RE})^{-1}\mathbf{G}_{RE}^H)\bm{\varsigma}_{r,\hat{k}}=\mathbf{P}\bm{\varsigma}_{r,\hat{k}}.\label{37}
\end{align}	
$\bm{\varsigma}_{r,\hat{k}}$ denotes an auxiliary variable.
Then, based on the leakage theory, 
the optimization problem with respect to $\bm{\varsigma}_{r,\hat{k}}$ can be expressed as
 \begin{subequations}\begin{align}
		\mathrm{P3}:\ &{\operatorname*{max}_{\bm{\varsigma}_{r,\hat{k}}}}\quad\bm{\varsigma}_{r,\hat{k}}^H\mathbf{P}^H\mathbf{g}_{r,\hat{k}}\mathbf{g}_{r,\hat{k}}^H\mathbf{P}\bm{\varsigma}_{r,\hat{k}}\\
		&\mathrm{s.t.}\quad\ \bm{\varsigma}_{r,\hat{k}}^H\mathbf{P}^H\mathbf{P}\bm{\varsigma}_{r,\hat{k}}=1.
	\end{align} \label{40}\end{subequations}
P3 is a generalized Rayleigh quotient (GRQ) problem, and the solution for $\bm{\varsigma}_{r,\hat{k}}$ is the eigenvector corresponding to the largest eigenvalue of $(\mathbf{P}^H\mathbf{P})^{-1}\mathbf{P}^H\mathbf{g}_{r,\hat{k}}\mathbf{g}_{r,\hat{k}}^H\mathbf{P}$. With ${\mathbf{u}}_{t,k}^*$ and $\hat{\mathbf{u}}_{t,\hat{k}}^*$, $\mathbb{A}$ can be obtained according to \eqref{34} and \eqref{35}. When $\sum_{\hat{k}=1}^{K_a}\hat{\mu}_{\hat{k}}+\sum_{k=1}^{{K}}{\mu}_k<1$, to maximize the SR, the transmit power needs to be fully utilized. The remaining power is allocated to the aerial Bob.

Based on the leakage theory, the sub‑optimization problem with respect to $\hat{\mathbf{w}}_{\hat{k}}$ can be represented as
 \begin{subequations}\begin{align}
\mathrm{P4}:\ &{\operatorname*{max}_{\hat{\mathbf{w}}_{\hat{k}}}}\quad\frac{\hat{\mathbf{w}}_{\hat{k}}^H\mathbf{H}_{BRS}^H\mathbf{H}_{BRS}\hat{\mathbf{w}}_{\hat{k}}}{\hat{\mathbf{w}}_{\hat{k}}^H(\hat{\mathbf{H}}_{BRS}^H\hat{\mathbf{H}}_{BRS}+\sigma_1^2\mathbf{I}_M)\hat{\mathbf{w}}_{\hat{k}}}\\
&\mathrm{s.t.}\quad\ \hat{\mathbf{w}}_{\hat{k}}^H\hat{\mathbf{w}}_{\hat{k}}=1.
\end{align} \label{41}\end{subequations}
 The solution for $\hat{\mathbf{w}}_{\hat{k}}$ is the unit eigenvector corresponding to the largest eigenvalue of $(\hat{\mathbf{H}}_{BRS}^H\hat{\mathbf{H}}_{BRS}+\sigma_1^2\mathbf{I}_M)^{-1}\mathbf{H}_{BRS}^H\mathbf{H}_{BRS}$, denoted as
 \begin{align}\hat{\mathbf{w}}_{\hat{k}}^*=\varepsilon_{\mathrm{max}}\{(\hat{\mathbf{H}}_{BRS}^H\hat{\mathbf{H}}_{BRS}+\sigma_1^2\mathbf{I}_M)^{-1}\mathbf{H}_{BRS}^H\mathbf{H}_{BRS}\}.\label{42}
 \end{align}
Similarly, the sub‑optimization problem with respect to ${\mathbf{w}}_{{k}}$ can be represented as
 \begin{subequations}\begin{align}
		\mathrm{P5}:\ &{\operatorname*{max}_{{\mathbf{w}}_{{k}}}}\quad\frac{{\mathbf{w}}_{{k}}^H\mathbf{G}_{BRS}^H\mathbf{G}_{BRS}{\mathbf{w}}_{{k}}}{{\mathbf{w}}_{{k}}^H(\hat{\mathbf{G}}_{BRS}^H\hat{\mathbf{G}}_{BRS}+\sigma_2^2\mathbf{I}_M){\mathbf{w}}_{{k}}}\\
		&\mathrm{s.t.}\quad\ {\mathbf{w}}_{{k}}^H{\mathbf{w}}_{{k}}=1.
	\end{align} \end{subequations}
 The solution for ${\mathbf{w}}_{{k}}$ can be denoted as
  \begin{align}{\mathbf{w}}_{{k}}^*=\varepsilon_{\mathrm{max}}\{(\hat{\mathbf{G}}_{BRS}^H\hat{\mathbf{G}}_{BRS}+\sigma_2^2\mathbf{I}_M)^{-1}\mathbf{G}_{BRS}^H\mathbf{G}_{BRS}\}.\label{44}
 \end{align}
 For the receive beamforming vectors, the minimum mean square error (MMSE) criterion can be adopted for derivation and calculation. 
 
 
Then, P2 can be described as a sub‑optimization problem with respect to $\mathbb{P}$, which can be expressed as
  \begin{subequations}
 	\begin{align}\mathrm{P6}:&{\operatorname*{max}_{\mathbb{P}}}\quad \frac{\sigma_6^2+\sigma_6^2\nu}{\sigma_6^2+(1-\mu_k)P_s\|({\tilde{\mathbf{u}}_{\hat{k}}^*})^H\mathbf{G}_{RE}^H\hat{\mathbf{u}}_{t,\hat{k}}^*\|^2}\label{46a}\\&\mathrm{s.t.}\ \eqref{s3}.
 	\end{align}
 \end{subequations}
 
 Physically, P6 seeks the optimal orientation combination of the rotatable antennas to maximize the SR. Due to the vast number of possible orientation combinations of the RAs within a single time slot, it is difficult to efficiently determine the optimal combination. To address this challenge, the auxiliary variables $\bm{\varpi}_1$ are introduced. Then, the optimization scheme for the RAs at the BS in the first time slot is
  \begin{subequations}\begin{align}
 		\mathrm{P7}:\ &\operatorname*{max}_{\bm{\varpi}_1}\ \frac{\bm{\varpi}_1^H\mathbf{A}_{2,\hat{k}}^H\mathbf{A}_{2,\hat{k}}\bm{\varpi}_1}{\bm{\varpi}_1^H(\mathbf{A}_{1,\hat{k}}^H\mathbf{A}_{1,\hat{k}}+\sigma_2^2\mathbf{I}_M)\bm{\varpi}_1}\\
 		&\mathrm{s.t.}\quad\ f_4(\theta_{\max})\leq|\bm{\varpi}_1|\leq1,\label{47b}
 	\end{align} \label{47}\end{subequations}
 where \begin{align}\mathbf{A}_{1,\hat{k}}[\bar{m}_1,:]&=\hat{\mathbf{H}}_{BRS}[\bar{m}_1,:]\odot(\hat{\mathbf{w}}_{\hat{k}}^*)^T,\\ \mathbf{A}_{2,\hat{k}}[\bar{m}_1,:]&={\mathbf{H}}_{BRS}[\bar{m}_1,:]\odot(\hat{\mathbf{w}}_{\hat{k}}^*)^T.\end{align} 
Similarly, the optimization of the RAs at the RS can be described as 
  \begin{subequations}\begin{align}
 	\!\!\!\!	\mathrm{P8}:\ &\operatorname*{max}_{{\bm{\varpi}}_2}\quad\frac{{\bm{\varpi}}_2^H\mathbf{H}_{BRS}\hat{\mathbf{w}}_{\hat{k}}^*(\hat{\mathbf{w}}_{\hat{k}}^*)^H\mathbf{H}_{BRS}^H{\bm{\varpi}}_2}{{\bm{\varpi}}_2^H(\hat{\mathbf{G}}_{BRS}{\mathbf{w}}_{{k}}^*({\mathbf{w}}_{{k}}^*)^H\hat{\mathbf{G}}_{BRS}^H+\sigma_1^2\mathbf{I}_M){\bm{\varpi}}_2}\\
 		&\mathrm{s.t.}\quad\ f_5(\theta_{\max})\leq|{\bm{\varpi}}_2|\leq1.
 	\end{align} \label{50}\end{subequations}
 In the second time slot, the RAs at the RS can be designed as
     \begin{subequations}
     	\begin{align}
  		\mathrm{P9}: &\operatorname*{max}_{{\bm{\varpi}}_3}{\bm{\varpi}}_3^H\mathbf{P}_{\hat{k}}^H\mathbf{a}_{3,\hat{k}}^H\mathbf{a}_{3,\hat{k}}\mathbf{P}_{\hat{k}}{\bm{\varpi}}_3\\
  		&\mathrm{s.t.}\quad\ f_6(\theta_{\max})\leq|\mathbf{P}_{\hat{k}}{\bm{\varpi}}_3|\leq1,
  	\end{align} \label{51}
  	\end{subequations}
 where 
 \begin{align}&\mathbf{a}_{3,\hat{k}} = \mathbf{g}_{r,\hat{k}}^T\odot(\hat{\mathbf{u}}_{t,\hat{k}}^*)^H,
\\ &\mathbf{A}_{3,\hat{k}}[:,e] = \text{conj}(\mathbf{G}_{RE}[:,e])\odot(\mathbf{P}\bm{\varsigma}_{r,\hat{k}}),\\
&\mathbf{P}_{\hat{k}} = \mathbf{I}_M - \mathbf{A}_{3,\hat{k}}(\mathbf{A}_{3,\hat{k}}^H\mathbf{A}_{3,\hat{k}})^{-1}\mathbf{A}_{3,\hat{k}}^H.
 	\end{align} 
 ${|\bm{\varpi}_1|}$ and ${|\bm{\varpi}_2|}$ denote the attenuation ratios of the RA gains at the BS and RS in the first time slot, respectively, which physically represent the orientation adjustments of the RAs to reduce interference. ${|\mathbf{P}_{\hat{k}}{\bm{\varpi}}_3|}$ denotes the attenuation ratio of the RA gain at the RS in the second time slot. $f_4(\theta_{\max})$, $f_5(\theta_{\max})$, and $f_6(\theta_{\max})$ denote the attenuation ratios associated with the maximum zenith angle $\theta_{\max}$, respectively. By relaxing the constraints, the solutions for P7, P8, and P9 can be obtained based on the GRQ. Specifically, the solutions for $\bm{\varpi}_1^*$, ${\bm{\varpi}}_2^*$, and ${\bm{\varpi}}_3^*$ can be expressed as 
     \begin{align}
 	&|{{\bm{\varpi}}}_1^*[m_2]|=\max\left\lbrace  \frac{{|{\hat{\bm{\varpi}}}_1^*[m_2]|}}{\|{{\hat{\bm{\varpi}}}_1^*}\|_{\infty}},f_4(\theta_{\max})\right\rbrace ,\label{58a}\\
 	&|{{{\bm{\varpi}}}}_2^*[\bar{m}_2]|=\max\left\lbrace \frac{{|{\hat{\bm{\varpi}}}_2^*[\bar{m}_2]|}}{\|{{\hat{\bm{\varpi}}}_2^*}\|_{\infty}}, f_5(\theta_{\max})\right\rbrace ,\label{59a}\\
 	&|{{\bm{\varpi}}}_3^*[\bar{m}_2]|=\max\left\lbrace \frac{{|{\mathbf{P}_{\hat{k}}[\bar{m}_2,:]\hat{\bm{\varpi}}}_3^*|}}{\|{{\mathbf{P}_{\hat{k}}\hat{\bm{\varpi}}}_3^*}\|_{\infty}}, f_6(\theta_{\max})\right\rbrace, \label{60a}
 \end{align} 
 where
    \begin{align}
 		&{\hat{\bm{\varpi}}}_1^* =\varepsilon_{\max}\{(\mathbf{A}_{1,\hat{k}}^H\mathbf{A}_{1,\hat{k}}+\sigma_2^2\mathbf{I}_M)^{-1}\mathbf{A}_{2,\hat{k}}^H\mathbf{A}_{2,\hat{k}}\},\\
 		&{{\hat{\bm{\varpi}}}}_2^*=\varepsilon_{\mathrm{max}}\{(\hat{\mathbf{G}}_{BRS}{\mathbf{w}}_{{k}}^*({\mathbf{w}}_{{k}}^*)^H\hat{\mathbf{G}}_{BRS}^H+\sigma_1^2\mathbf{I}_M)^{-1}\notag\\&\quad\qquad\qquad \mathbf{H}_{BRS}\hat{\mathbf{w}}_{\hat{k}}^*(\hat{\mathbf{w}}_{\hat{k}}^*)^H\mathbf{H}_{BRS}^H\},\\
 		&{\hat{\bm{\varpi}}}_3^*=\varepsilon_{\mathrm{max}}\{(\mathbf{P}_{\hat{k}}^H\mathbf{P}_{\hat{k}})^{-1}\mathbf{P}_{\hat{k}}^H\mathbf{a}_{3,\hat{k}}^H\mathbf{a}_{3,\hat{k}}\mathbf{P}_{\hat{k}}\}.
 	\end{align} 
  respectively. Accordingly, the directional gain will not exceed the adjustable range. \textcolor{black}{However, projecting the optimized antenna directional gain onto the physically realizable range may alter the optimized waveform.} To address this issue, the weight amplitudes of the corresponding RAs are reduced to maintain the waveform. The advantage of this adjustment is that it reduces power consumption. Specifically, the adjusted beamforming vectors can be expressed as 
      \begin{align}
  	&\hat{\mathbf{w}}_{\hat{k}}^{**}[m_2]= \begin{cases}\hat{\mathbf{w}}_{\hat k}^{*}[m_2]e^{j\angle{{{\bm{\varpi}}}_1^*[m_2]}},&\frac{{|{\hat{\bm{\varpi}}}_1^*[m_2]|}}{\|{{\hat{\bm{\varpi}}}_1^*}\|_{\infty}}\geq f_4,\\\frac{{|{\hat{\bm{\varpi}}}_1^*[m_2]|}\hat{\mathbf{w}}_{\hat k}^{*}[m_2]}{f_4\|{{\hat{\bm{\varpi}}}_1^*}\|_{\infty}}e^{j\angle{{{\bm{\varpi}}}_1^*[m_2]}},&\frac{{|{\hat{\bm{\varpi}}}_1^*[m_2]|}}{\|{{\hat{\bm{\varpi}}}_1^*}\|_{\infty}}<f_4,\end{cases}\label{58}\\
  	&\hat{\mathbf{u}}_{\hat{k}}^{**}[\bar{m}_2]= \begin{cases}\hat{\mathbf{u}}_{\hat{k}}^{*}[\bar{m}_2]e^{j\angle{{{\bm{\varpi}}}_2^*[\bar{m}_2]}},&\frac{{|{\hat{\bm{\varpi}}}_2^*[\bar{m}_2]|}}{\|{{\hat{\bm{\varpi}}}_2^*}\|_{\infty}}\geq f_5,\\\frac{{|{\hat{\bm{\varpi}}}_2^*[\bar{m}_2]|}\hat{\mathbf{u}}_{\hat{k}}^{*}[\bar{m}_2],}{f_5\|{{\hat{\bm{\varpi}}}_2^*}\|_{\infty}}e^{j\angle{{{\bm{\varpi}}}_2^*[\bar{m}_2]}}&\frac{{|{\hat{\bm{\varpi}}}_2^*[\bar{m}_2]|}}{\|{{\hat{\bm{\varpi}}}_2^*}\|_{\infty}}<f_5,\end{cases}\label{59}\\
  	&\hat{\mathbf{u}}_{t,\hat{k}}^{**}[\bar{m}_2]\notag\\&= \begin{cases}\hat{\mathbf{u}}_{t,\hat{k}}^{*}[\bar{m}_2]e^{j\angle{{{\bm{\varpi}}}_3^*[\bar{m}_2]}},&\frac{{|{\mathbf{P}_{\hat{k}}[\bar{m}_2,:]\hat{\bm{\varpi}}}_3^*|}}{\|{{\mathbf{P}_{\hat{k}}\hat{\bm{\varpi}}}_3^*}\|_{\infty}}\geq f_6,\\\frac{{|\mathbf{P}_{\hat{k}}[\bar{m}_2,:]{\hat{\bm{\varpi}}}_3^*|}\hat{\mathbf{u}}_{t,\hat{k}}^{*}[\bar{m}_2]}{f_6\|{{\mathbf{P}_{\hat{k}}\hat{\bm{\varpi}}}_3^*}\|_{\infty}}e^{j\angle{{{\bm{\varpi}}}_3^*[\bar{m}_2]}},&\frac{{|\mathbf{P}_{\hat{k}}[\bar{m}_2,:]{\hat{\bm{\varpi}}}_3^*|}}{\|{{\mathbf{P}_{\hat{k}}\hat{\bm{\varpi}}}_3^*}\|_{\infty}}<f_6.\end{cases}\label{60}
  \end{align} 
 According to \eqref{6}, the optimization of the zenith and azimuth angles of the RAs satisfies 
 \begin{align}
 	&\![(\vec{\mathbf{p}}_{m_2}^{BS})^T\vec{\mathbf{d}}_{m_2}^{\bar{m}_2}]_+^{2p}=	|{{\bm{\varpi}}}_1^*[m_2]|[(\vec{\mathbf{d}}_{m_2}^{1})^T\vec{\mathbf{d}}_{m_2}^{\bar{m}_2}]_+^{2p}=a_{1,m_2},\label{66a}\\
 	&\![(\vec{\mathbf{p}}_{\bar{m}_2}^{RS,1})^T\vec{\mathbf{d}}_{\bar{m}_2}^{m_2}]_+^{2p}=	|{{\bm{\varpi}}}_2^*[\bar{m}_2]|[(\vec{\mathbf{d}}^{\bar{m}_2}_{1})^T\vec{\mathbf{d}}^{m_2}_{\bar{m}_2}]_+^{2p}=a_{2,\bar{m}_2},\label{66b}\\
 	&\!	[(\vec{\mathbf{p}}_{\bar{m}_2}^{RS,2})^T\vec{\mathbf{k}}^{\hat{k}}_{\bar{m}_2}]_+^{2p}=	|{{\bm{\varpi}}}_3^*[\bar{m}_2]|[(\vec{\mathbf{k}}^{1}_{\bar{m}_2})^T\vec{\mathbf{k}}^{\hat{k}}_{\bar{m}_2}]_+^{2p}=a_{3,\bar{m}_2}.\label{66c}
 \end{align}
\textcolor{black}{Since the RA gain is direction‑dependent, the gains in undesired directions cannot maintain the same scaling ratio as those in the desired directions, leading to waveform mismatch.} By introducing auxiliary variables ${\bm{\varpi}}_4$, ${\bm{\varpi}}_5$, and ${\bm{\varpi}}_6$, the interference is minimized within the adjustable range. Then, under the constraints of \eqref{66a}, \eqref{66b}, and \eqref{66c}, the upper and lower bounds of the interference can be expressed as \eqref{67a}, \eqref{67b}, and \eqref{67c}, respectively. Here, $b_{m_2} = (\vec{\mathbf{d}}_{m_2}^{\bar{m}_1})^T\vec{\mathbf{d}}_{m_2}^{1}$, $c_{\bar{m}_2} = (\vec{\mathbf{d}}_{\bar{m}_2}^{m_1})^T\vec{\mathbf{d}}_{1}^{\bar{m}_2}$, and $d_{\bar{m}_2} = (\vec{\mathbf{k}}_{\bar{m}_2}^{e})^T\vec{\mathbf{k}}_{\bar{m}_2}^{1}$.
\begin{figure*}
 \begin{align}
	&\left[ a_{1,m_2}^{\frac{1}{2p}}b_{m_2}-\sqrt{(1-(a_{1,m_2}^{\frac{1}{2p}})^2)(1-b_{m_2}^2)}\right] _+^{2p}\leq\frac{{{\bm{\varpi}}}_4[m_2]B_{m_2,\bar{m}_1}^{BS}}{B_0}\leq\left[ a_{1,m_2}^{\frac{1}{2p}}b_{m_2}+\sqrt{(1-(a_{1,m_2}^{\frac{1}{2p}})^2)(1-b_{m_2}^2)}\right] _+^{2p},\label{67a}\\
	&\left[ a_{2,\bar{m}_2}^{\frac{1}{2p}}c_{\bar{m}_2}-\sqrt{(1-(a_{2,\bar{m}_2}^{\frac{1}{2p}})^2)(1-c_{\bar{m}_2}^2)}\right] _+^{2p}\leq\frac{{{\bm{\varpi}}}_5[\bar{m}_2]B_{m_1,\bar{m}_2}^{RS,1}}{B_0}\leq\left[ a_{2,\bar{m}_2}^{\frac{1}{2p}}c_{\bar{m}_2}+\sqrt{(1-(a_{2,\bar{m}_2}^{\frac{1}{2p}})^2)(1-c_{\bar{m}_2}^2)}\right] _+^{2p},\label{67b}\\
	&	\left[ a_{3,\bar{m}_2}^{\frac{1}{2p}}d_{\bar{m}_2}-\sqrt{(1-(a_{3,\bar{m}_2}^{\frac{1}{2p}})^2)(1-d_{\bar{m}_2}^2)}\right] _+^{2p}\leq\frac{{{\bm{\varpi}}}_6[\bar{m}_2]B_{\bar{m}_2,e}^{RS,2}}{B_0}\leq\left[ a_{3,\bar{m}_2}^{\frac{1}{2p}}d_{\bar{m}_2}+\sqrt{(1-(a_{3,\bar{m}_2}^{\frac{1}{2p}})^2)(1-d_{\bar{m}_2}^2)}\right] _+^{2p}.\label{67c}\\\hline\notag
\end{align}
\end{figure*}
Thus, the optimization schemes are as follows:
     \begin{subequations}
	\begin{align}
		\mathrm{P10}: &\operatorname*{min}_{{\bm{\varpi}}_4}{\bm{\varpi}}_4^T\tilde{\mathbf{A}}_{1,\hat{k}}^T\tilde{\mathbf{A}}_{1,\hat{k}}{\bm{\varpi}}_4\\
		&\mathrm{s.t.}\quad\ \eqref{67a},
	\end{align} \label{71a}
\end{subequations}
     \begin{subequations}
	\begin{align}
		\mathrm{P11}: &\operatorname*{min}_{{\bm{\varpi}}_5}{\bm{\varpi}}_5^T\tilde{\mathbf{A}}_{2,\hat{k}}^T\tilde{\mathbf{A}}_{2,\hat{k}}{\bm{\varpi}}_5\\
		&\mathrm{s.t.}\quad\ \eqref{67b},
	\end{align} \label{72a}
\end{subequations}
     \begin{subequations}
	\begin{align}
		\mathrm{P12}: &\operatorname*{min}_{{\bm{\varpi}}_6}{\bm{\varpi}}_6^T\tilde{\mathbf{A}}_{3,\hat{k}}^T\tilde{\mathbf{A}}_{3,\hat{k}}{\bm{\varpi}}_6\\
		&\mathrm{s.t.}\quad\ \eqref{67c},
	\end{align} \label{73a}
\end{subequations}
where
	\begin{align}\label{74a}
&\tilde{\mathbf{A}}_{1,\hat{k}} = \begin{bmatrix}
	\Re{(\hat{\mathbf{A}}_{1,\hat{k}})}\\ \Im{(\hat{\mathbf{A}}_{1,\hat{k}})}
\end{bmatrix},\\
&\tilde{\mathbf{A}}_{2,\hat{k}} = \begin{bmatrix}
	\Re{(({\mathbf{w}}_{{k}}^*)^H\hat{\mathbf{G}}_{BRS}^H)}\\ \Im{(({\mathbf{w}}_{{k}}^*)^H\hat{\mathbf{G}}_{BRS}^H)}
\end{bmatrix},\\
&\tilde{\mathbf{A}}_{3,\hat{k}} = \begin{bmatrix}
	\Re{(\hat{\mathbf{A}}_{3,\hat{k}})}\\ \Im{(\hat{\mathbf{A}}_{3,\hat{k}})}
\end{bmatrix},\\
&\hat{\mathbf{A}}_{1,\hat{k}}[\bar{m}_1,:] = \hat{\mathbf{H}}_{BRS}[\bar{m}_1,:]\odot(\hat{\mathbf{w}}_{\hat{k}}^{**})^T,\\
&\hat{\mathbf{A}}_{3,\hat{k}}[e,:] = (\mathbf{G}_{RE}[:,e])^H\odot(\hat{\mathbf{u}}_{t,\hat{k}}^{**})^T.
\end{align} 
P10, P11, and P12 can be solved using the CVX toolbox. With the gain patterns in both the desired and undesired directions available, the orientations of all RAs can be determined. Given optimized variables, $\rho^*$ can be obtained using the quadratic formula. 
\begin{align}\rho^*=\frac{-B+\sqrt{B^2-4AC}}{2A},\label{49}
\end{align}
where $A = (a_1a_4-a_2a_3)P_0^2$, $B = (\sigma_2^2a_1+\sigma_1^2a_3)P_0+2a_2a_3P_0^2$, $C = -a_2a_3P_0^2-\sigma_1^2a_3P_0$, $a_1 = |(\hat{\mathbf{u}}_{\hat{k}}^{**})^H\mathbf{H}_{BRS}\hat{\mathbf{w}}^{**}_{\hat{k}}|^2$, $a_2 = |(\hat{\mathbf{u}}_{\hat{k}}^{**})^H{\hat{\mathbf{G}}}_{BRS}{\mathbf{w}}_{{k}}^{**}|^2$, $a_3 = | (\mathbf{u}_k^{**})^H\mathbf{G}_{BRS}{\mathbf{w}_k^{**}}|^2$, and $a_4 = |(\mathbf{u}_k^{**})^H\hat{\mathbf{H}}_{BRS}\hat{\mathbf{w}}^{**}_{\hat{k}}|^2$. Correspondingly, $\mu_k$ can be obtained using the quadratic formula.
 For clarity, the detailed steps are illustrated in Algorithm \ref{alg1}.
 
\textcolor{black}{ It is worth noting that Algorithm \ref{alg1} is a finite-step sequential optimization procedure rather than an iterative alternating optimization algorithm. The GRQ-based subproblems are directly solved by the eigenvectors associated with the largest generalized eigenvalues, while the constrained subproblems are solved using CVX and the power-allocation factors are obtained in closed form. Hence, no outer iteration is required and Algorithm \ref{alg1} terminates after a finite number of optimization steps. Since the original problem is non-convex and several variables are optimized sequentially with relaxed constraints, global optimality of the original problem is not guaranteed.}
 
 \begin{algorithm}
 	\renewcommand{\algorithmicrequire}{\textbf{Input:}}
 	\renewcommand{\algorithmicensure}{\textbf{Output:}}
 	\caption{Proposed GRQ-Based Step-by-Step Optimization Algorithm}\label{alg1}
 	\begin{algorithmic}[1]
 		\REQUIRE $P_0$, $P_s$, $M$, $\{\sigma_{i}\}_{i=1}^{6}$.
 		\STATE Initialization: $\zeta_k=1$, $\vec{\mathbf{d}}_{m_2}^{1}=[x_{m_2}^{1},y_{m_2}^{1},z_{m_2}^{1}]^T$, $\vec{\mathbf{d}}_{1}^{\bar{m}_2}=[x^{\bar{m}_2}_{1},y^{\bar{m}_2}_{1},z^{\bar{m}_2}_{1}]^T$, and $\vec{\mathbf{k}}_{\bar{m}_2}^{1}=[x_{\bar{m}_2}^{1},y_{\bar{m}_2}^{1},z_{\bar{m}_2}^{1}]^T$.
 		\STATE Compute the beamforming vector $\mathbf{u}_{t,k}^*$ for the RS using MRT.
 		\STATE Optimize the beamforming vector $\hat{\mathbf{u}}_{t,\hat{k}}^*$ using the NSP and GRQ method according to \eqref{37} and \eqref{40}.
 		\STATE Compute $\hat{\mathbf{w}}_{\hat{k}}^*$ and $\mathbf{w}_k^*$ according to \eqref{42} and \eqref{44}, respectively.
 		 \STATE Compute the receive beamforming matrices $\mathbf{U}^*$ and $\hat {\mathbf{U}}^*$ using the MMSE criterion.
 		\STATE Compute $\bm{\varpi}_1^*$, $\bm{\varpi}_2^*$, and $\bm{\varpi}_3^*$ according to \eqref{58a}, \eqref{59a}, and \eqref{60a}, respectively.
 		\STATE Update beamforming vectors $\hat{\mathbf{w}}_{\hat{k}}^{**}$, $\hat{\mathbf{u}}_{\hat{k}}^{**}$, and $\hat{\mathbf{u}}_{t,\hat{k}}^{**}$ according to \eqref{58},  \eqref{59}, and \eqref{60}.
 		 \STATE Solve P10, P11, and P12 using the CVX toolbox to obtain $\bm{\varpi}_4^*$, $\bm{\varpi}_5^*$, and $\bm{\varpi}_6^*$, respectively.
 		 \STATE Given $\{\bm{\varpi}_i^*\}_{i=1}^6$, $\mathbb{P}^*$ can be obtained.
 		  \STATE  Compute $\rho^*$ and $\mu_k^*$ using the quadratic formula.
 		\ENSURE $\mathbb{P}^*$, $\mu_k^*$, $\mathbf{w}_k^*$, $\hat{\mathbf{w}}_{\hat{k}}^{**}$, $\mathbf{u}_{t,k}^*$, $\hat{\mathbf{u}}_{t,\hat{k}}^{**}$, $\mathbf{U}^*$, $\mathbf{\hat{U}}^{**}$, $\tilde{\mathbf{u}}_{\hat{k}}^*$.
 	\end{algorithmic}
 \end{algorithm}
\section{Learning-Based Multi-User Case}\label{sec:4}
In this section, a learning-based solution for multi-Bob scenarios is investigated. Distributional Soft Actor-Critic (DSAC) with Three refinements (DSAC-T or DSACv2) is a distributional reinforcement learning algorithm designed for high-dimensional continuous control tasks~\cite{Duan2025}. Compared to DSAC~\cite{Duan2022}, its three refinements are: expectation replacement, twin value distribution learning, and variance-based critic gradient adjustment. It has been validated that DSAC-T achieves superior policy performance, stability, and $Q$-value estimation accuracy compared to mainstream reinforcement learning methods such as Soft Actor-Critic (SAC). However, in scenarios with extremely high-dimensional action spaces, the method still faces challenges such as an excessively large policy search space and slow convergence. In this section, we attempt to address this challenge by solving a subset of variables using traditional convex optimization algorithms to reduce the dimensionality of the action space. By combining convex optimization with DSAC-T, the interpretability of the overall approach is enhanced.
\subsection{Beamforming Scheme}
For multi‑Bob scenarios, the key lies in simultaneously reducing inter‑user interference and improving the secrecy rate under a limited transmit power constraint.

With the orientations of the antennas fixed, the transmission scheme for the first time slot can be determined by solving P4 and P5. The MRT transmission scheme may cause severe multi‑user interference and is therefore no longer applicable. Fortunately, compared with conventional designs that exploit parameters such as the number of antennas and transmit power, adjusting gain patterns for all antennas offers a new dimension for beam pattern design. Moreover, the RIS reflection path can be effectively exploited to further reduce interference.

The role of the RIS is to reflect different signals to their corresponding Bob locations, thereby concentrating the signal power at the Bob positions. Then, the inter‑user interference is reduced by optimizing the transmit weights. Consequently, for the RIS phase shift matrix $\mathbf{\Theta}$, we have
   \begin{subequations}\begin{align}
		\mathrm{P13}: &\operatorname*{max}_{\bm{\varrho}}\frac{\bm{\varrho}^H\sum_{k=1}^{K}\mathbf{A}_{4,k}\bm{\varrho}}{\bm{\varrho}^H\sum_{k=1}^{K}\mathbf{A}_{5,k}\bm{\varrho}}\\
		&\color{black}{\mathrm{s.t.}\ \alpha_q\in(1,\alpha_{\mathrm{max}}],\forall q.}
	\end{align} \label{80}\end{subequations}
where
\begin{align}&\mathbf{A}_{4,k}= (1-\zeta_k)(\mathrm{diag}(\mathbf{f}_{R,k}))^H\mathbf{H}_{BRIS}\mathbf{H}_{BRIS}^H\mathrm{diag}(\mathbf{f}_{R,k}),\\
			&\mathbf{A}_{5,k}=\sigma_4^2(1-\zeta_k)(\mathrm{diag}(\mathbf{f}_{R,k}))^H\mathrm{diag}(\mathbf{f}_{R,k}),\\&\mathbf{\Theta}\triangleq(\mathrm{diag}(\bm{\varrho}))^H.\end{align} 
\textcolor{black}{
		Due to the amplitude constraints, P13 cannot be directly solved as a standard GRQ problem. We first relax these constraints to determine the optimal direction of the RIS coefficient vector. Since the GRQ is invariant to any nonzero common scaling, the resulting generalized eigenvector can be scaled without changing the objective value. Accordingly, the largest common scaling factor satisfying the maximum reflection amplitude is adopted, yielding
\begin{align}\bar{\bm{\varrho}}=\frac{\alpha_{\mathrm{max}}\varepsilon_{\mathrm{max}}\{(\sum_{k=1}^{K}\mathbf{A}_{5,k})^{-1}\sum_{k=1}^{K}\mathbf{A}_{4,k}\}}{\|\varepsilon_{\mathrm{max}}\{(\sum_{k=1}^{K}\mathbf{A}_{5,k})^{-1}\sum_{k=1}^{K}\mathbf{A}_{4,k}\}\|_{\infty}}.\label{85}
\end{align}
Since \eqref{85} may not satisfy the minimum reflection amplitude requirement, the following auxiliary variables are introduced:  
\begin{align}&\bm{\varrho}_{\mathrm{aux}}=\begin{bmatrix}\bm{\varrho}_1\\ \bm{\varrho}_2\end{bmatrix},\\&\sum_{k=1}^{K}\mathbf{A}_{4,k} = \begin{bmatrix}
\mathbf{A}_{1,1}\ \mathbf{A}_{1,2}\\\mathbf{A}_{2,1}\ \mathbf{A}_{2,2}
\end{bmatrix} ,\\&\sum_{k=1}^{K}\mathbf{A}_{5,k} = \begin{bmatrix}
\mathbf{B}_{1,1}\ \mathbf{B}_{1,2}\\\mathbf{B}_{2,1}\ \mathbf{B}_{2,2}
\end{bmatrix}, \end{align}
$\bm{\varrho}_1$ and $\bm{\varrho}_2$ denote the sets of RIS elements that satisfy and do not satisfy the constraint, respectively. 
With $\bm{\varrho}_1$ fixed and by relaxing the constraint, we have
\begin{align}
		\mathrm{P14}: &\operatorname*{max}_{\tilde{\bm{\varrho}}_2}\frac{\tilde{\bm{\varrho}}_2^H\tilde{\mathbf{A}}_1\tilde{\bm{\varrho}}_2}{\tilde{\bm{\varrho}}_2^H\tilde{\mathbf{A}}_2 \tilde{\bm{\varrho}}_2}.\label{87new}
	\end{align} 
where \begin{align}&\tilde{\bm{\varrho}}_2=\begin{bmatrix}1\\{\bm{\varrho}}_2\end{bmatrix},\\&\tilde{\mathbf{A}}_1 =\begin{bmatrix}
		\bm{\varrho}_1^H\mathbf{A}_{1,1}\bm{\varrho}_1\ \bm{\varrho}_1^H\mathbf{A}_{1,2}\\\mathbf{A}_{2,1}\bm{\varrho}_1\ \mathbf{A}_{2,2}
	\end{bmatrix},\\&\tilde{\mathbf{A}}_2 =\begin{bmatrix}
	\bm{\varrho}_1^H\mathbf{B}_{1,1}\bm{\varrho}_1\ \bm{\varrho}_1^H\mathbf{B}_{1,2}\\\mathbf{B}_{2,1}\bm{\varrho}_1\ \mathbf{B}_{2,2}
	\end{bmatrix}.\end{align}  The solution to P14 is $\tilde{\bm{\varrho}}_2^*= \frac{\varepsilon_{\mathrm{max}}\{\tilde{\mathbf{A}}_2^{-1}\tilde{\mathbf{A}}_1\}}{x_0^*}$, where $x_0^*$ denotes the first element of $\varepsilon_{\mathrm{max}}\{\tilde{\mathbf{A}}_2^{-1}\tilde{\mathbf{A}}_1\}$. Then, by projecting 
${\bm{\varrho}}_2^*$ onto $(1,\alpha_{\mathrm{max}}]$, $\mathbf{\Theta}^*$ can be obtained.}

For the transmit beamforming vector ${\mathbf{v}}_{t,{k}}$, we have 
   \begin{subequations}\begin{align}
		\color{black}\mathrm{P15}: &\operatorname*{max}_{{\mathbf{v}}_{t,k}}\frac{{\mathbf{v}}_{t,k}^H\mathbf{A}_{6,k}{\mathbf{v}}_{t,k}}{{\mathbf{v}}_{t,k}^H\mathbf{A}_{7,k}{\mathbf{v}}_{t,k}}\\
		&\mathrm{s.t.}\ {\mathbf{v}}_{t,k}^H{\mathbf{v}}_{t,k}=1.
	\end{align} \label{84}\end{subequations}
where \begin{align}\mathbf{A}_{6,k}&= (1-\zeta_k)\mathbf{H}_{BRIS}^H(\mathbf{\Theta}^*)^H\mathbf{f}_{R,k}\mathbf{f}_{R,k}^H\mathbf{\Theta}^*\mathbf{H}_{BRIS},\\\mathbf{A}_{7,k}&=\sum_{i=1,i\neq k}^{K}(1-\zeta_i)\mathbf{H}_{BRIS}^H(\mathbf{\Theta}^*)^H\mathbf{f}_{R,i}\mathbf{f}_{R,i}^H\mathbf{\Theta}^*\mathbf{H}_{BRIS}\notag\\&\quad+\sigma_3^2\mathbf{I}_M.\end{align}
The solution for $\mathbf{v}_{t,k}$ is the unit eigenvector corresponding to the largest eigenvalue of $\mathbf{A}_{7,k}^{-1}\mathbf{A}_{6,k}$, denoted as
\begin{align}\mathbf{v}_{t,k}^*=\varepsilon_{\mathrm{max}}\{\mathbf{A}_{7,k}^{-1}\mathbf{A}_{6,k}\}.\label{89}
\end{align}

Correspondingly, the optimization scheme from the RS to the ground Bobs can be expressed as
   \begin{subequations}\begin{align}
		\color{black}\mathrm{P16}: &\operatorname*{max}_{\mathbf{u}_{t,k}}\frac{\zeta_k{\mathbf{u}}_{t,k}^H\mathbf{f}_{s,k}\mathbf{f}_{s,k}^H{\mathbf{u}}_{t,k}}{{\mathbf{u}}_{t,k}^H(\sum_{i=1,i\neq k}^{K}\zeta_i\mathbf{f}_{s,i}\mathbf{f}_{s,i}^H+\sigma_3^2\mathbf{I}_M){\mathbf{u}}_{t,k}}\\
		&\mathrm{s.t.}\ \mathbf{u}_{t,k}^H\mathbf{u}_{t,k}=1.
	\end{align} \label{67}\end{subequations}
The solution for $\mathbf{u}_{t,k}$ is the unit eigenvector corresponding to the largest eigenvalue of $(\sum_{i=1,i\neq k}^{K}\zeta_i\mathbf{f}_{s,i}\mathbf{f}_{s,i}^H+\sigma_3^2\mathbf{I}_M)^{-1}\mathbf{f}_{s,k}\mathbf{f}_{s,k}^H$, denoted as
\begin{align}\mathbf{u}_{t,k}^*=\zeta_k\varepsilon_{\mathrm{max}}\{(\sum_{i=1,i\neq k}^{K}\zeta_i\mathbf{f}_{s,i}\mathbf{f}_{s,i}^H+\sigma_3^2\mathbf{I}_M)^{-1}\mathbf{f}_{s,k}\mathbf{f}_{s,k}^H\}.\label{91}
\end{align}

For the communication from the RS to the aerial Bobs, null‑space projection and leakage theory are exploited to maximize the SR. Similar to \eqref{37}, by introducing an auxiliary variable $\bm{\varrho}_{r,\hat{k}}$ and setting $\hat{\mathbf{u}}_{t,\hat{k}}^*=\mathbf{P}\bm{\varrho}_{r,\hat{k}}$, we obtain 
 \begin{subequations}\begin{align}
		\color{black}\mathrm{P17}:\ &{\operatorname*{max}_{\bm{\varrho}_{r,\hat{k}}}}\quad\frac{\bm{\varrho}_{r,\hat{k}}^H\mathbf{P}^H\mathbf{g}_{r,\hat{k}}\mathbf{g}_{r,\hat{k}}^H\mathbf{P}\bm{\varrho}_{r,\hat{k}}}{\bm{\varrho}_{r,\hat{k}}^H\mathbf{P}^H(\sum_{i=1,i\neq \hat{k}}^{K_a}\mathbf{g}_{r,i}\mathbf{g}_{r,i}^H+\sigma_5^2\mathbf{I}_M)\mathbf{P}\bm{\varrho}_{r,\hat{k}}}\\
		&\mathrm{s.t.}\quad\ \bm{\varrho}_{r,\hat{k}}^H\bm{\varrho}_{r,\hat{k}}=1.
	\end{align} \label{69}\end{subequations}
The solution for $\bm{\varrho}_{r,\hat{k}}$ is the eigenvector corresponding to the largest eigenvalue of $\mathbf{A}_{8,\hat{k}}=(\mathbf{P}^H(\sum_{i=1,i\neq \hat{k}}^{K_a}\mathbf{g}_{r,i}\mathbf{g}_{r,i}^H+\sigma_5^2\mathbf{I}_M)\mathbf{P})^{-1}\mathbf{P}^H\mathbf{g}_{r,\hat{k}}\mathbf{g}_{r,\hat{k}}^H\mathbf{P}$. Then, we have
\begin{align}\hat{\mathbf{u}}_{t,\hat{k}}^*=\frac{\mathbf{P}\varepsilon_{\max}\{\mathbf{A}_{8,\hat{k}}\}}{\|\mathbf{P}\varepsilon_{\max}\{\mathbf{A}_{8,\hat{k}}\}\|}.\label{93}\end{align}
Then, the pointing vectors of the RAs in the second time slot need to be optimized to improve the SR. For the ground Bobs, there are $2^K$ possible assignments of the communication path. Correspondingly, power allocation needs to be performed based on the selected communication paths. To address these challenges, a learning‑based approach is employed.

\subsection{Optimization of \textcolor{black}{Pointing} Vectors and Path Selection}
A DSAC‑T method for hybrid discrete and continuous actions is proposed to address the optimal communication path selection for ground Bobs and the RA orientation for aerial Bobs.
\subsubsection{Mapping Relationships}
 Firstly, the mapping relationships among the state space, action space, reward, and environment are introduced, as detailed below.
\begin{itemize}
	\item   Action space:  A probability matrix $\bm{\Upsilon}^i$ is introduced to represent the probability with which each Bob selects each path, which is denoted as
	\begin{equation}
		\bm{\Upsilon}^i=[\mathbf{a}_1^i,\mathbf{a}_2^i,\ldots,\mathbf{a}_K^i]^T,
	\end{equation}
	where $\mathbf{a}_k^i=[p_{k,1}^i,p_{k,2}^i]^T$ denotes the probability that the $k$-th ground Bob selects each path, $p_{k,1}^i+p_{k,2}^i=1$. For example, $p_{k,1}^i$ denotes the probability of selecting the RS‑forwarding path, and $p_{k,2}^i$ denotes the probability of selecting the RIS‑assisted communication path. For the orientations of the RAs in the second slot, the zenith and azimuth angles of the antennas are regarded as the actions, i.e., ${\bm{\theta}_{\bar{m}}^{RS,2}}^i=[{\theta_{1}^{RS,2}}^i,{\theta_{2}^{RS,2}}^i,\ldots,{\theta_{M}^{RS,2}}^i]^T$ and ${\bm{\vartheta}_{\bar{m}}^{RS,2}}^i=[{\vartheta_{1}^{RS,2}}^i,{\vartheta_{2}^{RS,2}}^i,\ldots,{\vartheta_{M}^{RS,2}}^i]^T$. Then, $K+2M$ actions can be obtained. Thus, the action set at the $i$-th step can be denoted as
		\textcolor{black}{\begin{equation}
		a^i=\{\bm{\Upsilon}^i[:,1],{\bm{\theta}_{\bar{m}}^{RS,2}}^i,{\bm{\vartheta}_{\bar{m}}^{RS,2}}^i\}.
	\end{equation}}
	\item  State space: 
	Under different actions, the power allocation factors and transmission rates corresponding to different users may vary. Therefore, the state space is denoted as 
	\begin{equation}
		s^i=\{{\mathbb{A}^i},\bm{\gamma}^i,\hat{\bm{\gamma}}^i\}.
	\end{equation}
	where $\bm{\gamma}^i = [{\gamma}_{1}^i,\ldots,{\gamma}_{K}^i]^T$ and $\hat{\bm{\gamma}}^i= [\hat{\gamma}_{1}^i,\ldots,\hat{\gamma}_{K_a}^i]^T$.
	\item   Reward: To ensure learning stability, the reward is designed as
	\begin{equation}\color{black}
		r^{i}=R_{s}^{i}-\delta(\sum_{k=1}^{K}[\nu-\gamma_k^i]^++\sum_{\hat{k}=1}^{K_a}[{\nu}-\hat{\gamma}_{\hat k}^i]^+),\label{97}
	\end{equation}
	where $\delta$ denotes the penalty factor. 
\end{itemize}
\subsubsection{ Critic update}
A self-consistency operator $\mathcal{T}^{\hat{\pi}}$ is applied to learn the value distribution, which can be expressed as
\begin{equation}
	\mathcal{T}^{\hat{\pi}}\mathcal{Z}(s,a)=r+\omega(\mathcal{Z}(s^{\prime},a^{\prime})-\alpha\log\hat{\pi}(s^{\prime}|a^{\prime})),
\end{equation}
where $\omega$ and $\alpha$ denote the discount factor and  the temperature
coefficient, respectively. $\mathcal{Z}(s,a)$ can be modeled as a Gaussian distribution with mean equal to the $Q$ function. To make the predicted distribution approximate the Bellman target distribution, the new value distribution $\mathcal{Z}^*$ can be obtained by minimizing the Kullback-Leibler divergence, i.e., 
\begin{equation}
	\mathcal{Z}^*=\mathrm{arg}\underset{\mathcal{Z}}{\min}\underset{(s,a)\sim\rho_{\hat{\pi}}}{\mathbb{E}}[D_{\mathrm{KL}}(\mathcal{T}^{\hat{\pi}}\mathcal{Z}^{\prime}(\cdot|s,a),\mathcal{Z}(\cdot|s,a))],\label{77}
\end{equation}
where $D_{\mathrm{KL}}$ denotes the Kullback-Leibler divergence.

Based on \eqref{77}, a policy evaluation scheme can be obtained. Using the action mean and variance output by the actor network, a Gaussian distribution can be constructed, from which the probability of each ground Bob selecting the RS‑forwarding path is sampled. After executing the actions, the current reward and the next state can be obtained. Then, the experience $(a,s,r,s^{\prime})$ is stored in the replay buffer $\mathcal{B}$. Once the replay buffer $\mathcal{B}$ has collected sufficient experiences, a batch of ${B}_s$ experiences is sampled. The policy network is then updated by minimizing the following function: 
\begin{equation}
	\begin{aligned}&J_{\mathcal{Z}}(\hat{\theta})=-\underset{(a,s,r,s^{\prime})\sim\mathcal{B}}{\mathbb{E}}\left[\log\Xi(y_{z}|\mathcal{Z}_{\hat{\theta}}(\cdot|s,a))]\right],
	\end{aligned}
\end{equation}
where
\begin{equation}
	y_z=r+\omega(\mathcal{Z}(s^{\prime},a^{\prime})-\alpha\log\hat{\pi}_{\hat{\phi}}(s^{\prime}|a^{\prime})).\label{101}
\end{equation}
$\hat{\theta}$ and $\hat{\phi}$ denote the parameters of the actor network and the critic network, respectively. Accordingly, the target value after substituting the expected value is expressed as 
\begin{equation}
	y_q=r+\omega(Q_{\hat{\theta}}(s^{\prime},a^{\prime})-\alpha\log\hat{\pi}_{\hat{\phi}}(s^{\prime}|a^{\prime})).\label{102}
\end{equation}
Using twin value distribution learning, we have $y_{z}^{\min}$ and $y_{q}^{\min}$ representing the value with the minimum mean. Then, the critic update gradient can be denoted as
\begin{equation}\begin{aligned}&\nabla_{\hat{\theta}_i}J_{\mathcal{Z}}(\hat{\theta}_i)\approx\mathbb{E}\Big[-\frac{(y_{q}^{\min}-Q_{\hat{\theta}_i}(s,a))}{\sigma_{\hat{\theta}_i}(s,a)^{2}}\nabla_{\hat{\theta}_i}Q_{\hat{\theta}_i}(s,a)\\&-\frac{(C(y_{z}^{\min};\bar{b})-Q_{\hat{\theta}_i}(s,a))^{2}-\sigma_{\hat{\theta}_i}(s,a)^{2}}{\sigma_{\hat{\theta}_i}(s,a)^{3}}\nabla_{\hat{\theta}_i}\sigma_{\hat{\theta}_i}(s,a)\Big],\end{aligned}\label{103}\end{equation}
where 
\begin{equation}
	C(y_{z}^{\min};\bar{b})\triangleq\operatorname{clip}\left(y_{z}^{\min},Q_{\hat{\theta}}(s,a)-\bar{b},Q_{\hat{\theta}}(s,a)+\bar{b}\right)
\end{equation}
denotes the clipping function. $\bar{b}$ denotes the adaptive clipping boundary.
To reduce the fluctuation caused by $\sigma_{\hat{\theta}_i}(s,a)$ on $\nabla_{\hat{\theta}_i}J_{\mathcal{Z}}(\hat{\theta}_i)$, $\tau_i=\mathbb{E}_{(s,a)\sim\mathcal{B}}\begin{bmatrix}\sigma_{\hat{\theta}_i}(s,a)^2\end{bmatrix}$ is introduced. Then, the gradient is updated as 
\begin{equation}\begin{aligned}&\nabla_{\hat{\theta}_{i}}J_{\mathcal{Z}}^{*}(\hat{\theta}_{i})\approx\\&(\tau_{i}+\epsilon_{1})\mathbb{E}\Big[-\frac{\left(y_{q}^{\min}-Q_{\hat{\theta}_{i}}(s,a)\right)}{\sigma_{\hat{\theta}_{i}}(s,a)^{2}+\epsilon}\nabla_{\hat{\theta}_{i}}Q_{\hat{\theta}_{i}}(s,a)\\&-\frac{\left(C(y_{z}^{\min},\bar{b}_{i})-Q_{\hat{\theta}_{i}}(s,a)\right)^{2}-\sigma_{\hat{\theta}_{i}}(s,a)^{2}}{\sigma_{\hat{\theta}_{i}}(s,a)^{3}+\epsilon_2}\nabla_{\hat{\theta}_{i}}\sigma_{\hat{\theta}_{i}}(s,a)\Big],\end{aligned}\label{105}\end{equation}
where $\epsilon_1$ and $\epsilon_{2}$ denote small positive constants.
\subsubsection{ Actor update}
The goal is to find a better policy for action selection. The entropy-augmented objective is maximized to update the policy. The actor objective can be represented as
\begin{equation}J_{\pi}(\hat\phi)=\underset{s\sim\mathcal{B},a\sim\pi_{\hat\phi}}{\mathbb{E}}[\underset{i=1,2}{\min}Q_{\hat{\theta}_i}(s,a)-\alpha\log(\pi_{\hat\phi}(a|s))].\label{106}\end{equation}
\textcolor{black}{For clarity, the DSAC‑T‑based scheme is illustrated in Algorithm \ref{alg2}.}
 \begin{algorithm}
	\color{black}
	\renewcommand{\algorithmicrequire}{\textbf{Input:}}
	\renewcommand{\algorithmicensure}{\textbf{Output:}}
	\caption{\textcolor{black}{Proposed DSAC-T-Based Optimization Algorithm}}\label{alg2}
	\begin{algorithmic}[1]
	\REQUIRE $P_0$, $P_s$, $M$, $K$, $K_a$, $\theta_{\max}$, $\{\sigma_{i}\}_{i=1}^{6}$, penalty factor, discount factor,
	temperature coefficient, replay-buffer size, batch size, and maximum number of episodes.
	\STATE Initialize the actor and twin critic network parameters and the replay buffer.
		\FOR{$\mathrm{episode} = 1,\ldots, \mathcal{I}_e$}
\STATE Initialize the environment and obtain the initial state.
\FOR{$i = 1, \ldots, \mathcal{I}$}
\STATE Compute the beamforming vectors.
\STATE Construct the state space $s^{i}$.
\STATE Generate the action distribution using the actor network.
\STATE Obtain the path-selection probabilities
$\boldsymbol{\Upsilon}^{i}$ and the RA pointing angles.
\STATE Construct the action space
$a^i$.
\STATE Calculate $r^{i}$ according to \eqref{97}.
\STATE Observe $\mathbf{s}^{i+1}$ and store
$(s^{i},a^{i},r^{i},s^{i+1})$
in $\mathcal{B}$.
\IF{$|\mathcal{B}| \geq B_s$}
\STATE Sample $B_s$ experiences from $\mathcal{B}$.
\STATE Generate the next action
$a^{i+1}$.
\STATE Compute $y_z$ and $y_q$ according to
\eqref{101} and \eqref{102}.
\STATE Update the twin critics according to \eqref{103}--\eqref{105}.
\STATE Update the actor network according to \eqref{106}.
\ENDIF
\ENDFOR
\ENDFOR
\ENSURE  $\mathbb{A}^*$, $\mathbb{B}^*$, $\mathbb{P}^*$, $\mathbb{U}^*$, $\mathbb{V}^*$, $\mathbb{W}^*$.
\end{algorithmic}
\end{algorithm}

\section{Simulation Results}\label{sec:5}
In this section, simulation results are presented to evaluate the performance of the relay network with hybrid arrays. Similar to~\cite{Zheng2026}, the system operates at a frequency of 2.4 GHz, and the noise power at all receivers is assumed to be -80 dBm. The antenna spacing is set to half a wavelength, and the antenna size is $D=\frac{\lambda^2}{4\pi}$. Unless otherwise specified, the transmit powers at the BS and the relay are $P_0 = 20$ dBm and $P_s = 20$ dBm, respectively. The number of antennas per array is $M=36$. The numbers of aerial and ground users are both 10. The maximum zenith angle is $\theta_{\max}=\frac{2\pi}{5}$. The antenna directivity factor is set to $p=2$. The number of RIS elements and the maximum reflection amplitude are $Q=64$ and $\alpha_{\mathrm{max}}=10$, respectively. The position vectors of the BS, RS, and RIS center are $[-30, 0, 15]^T$,  $[-30, 25, 12]^T$, and $[-15, 15, 8]^T$, respectively, where all coordinates are in meters. Ground Bobs are randomly distributed between the RS and the BS at a height of 1.5 m. The coordinates of aerial Bobs are randomly generated within the ranges of [-50, -40] for the $x$‑axis, [20, 25] for the $y$‑axis, and [15, 20] for the $z$‑axis. 
\textcolor{black}{The hyperparameters of the learning-based algorithm in the multi-user case are listed in Table~\ref{lab1}. }
\textcolor{black}{The following six baseline schemes are used for comparison:}
\begin{itemize}
	\item \textbf{Fixed orientation:} The orientation of each RA remains at its initial setting, i.e., pointing toward the desired reference target. RS with fixed orientation indicates that the orientation of each RA at the RS in the second time slot remains at its initial setting.
	\item \textbf{Equal power allocation:} In the \textcolor{black}{single ground–aerial user pair} scenario, equal power allocation is adopted at the BS and RS, i.e., $\rho=0.5$ and $\mu_k=0.5$.
	\item \textbf{Isotropic antennas:} The BS with isotropic antennas means that all antennas at the BS are isotropic, i.e., $p=0$. The BS and RS with isotropic antennas means that all antennas at both the BS and the RS are isotropic.
	\item \textbf{Random orientation:} The orientations of the RAs at the RS in the second time slot are randomly generated within the adjustable range.
	\item \textbf{Random path selection:} The communication paths for the ground Bobs are randomly selected from either the RIS reflection path or the RS path.
	\item \textcolor{black}{\textbf{RS‑only transmission:} In the second time slot, transmission is performed solely via the RS. Corresponding RS‑only transmission cases are designed under Algorithm \ref{alg2}, isotropic antennas, and fixed orientations, respectively.}
\end{itemize}	
The proposed Algorithm 1 and Algorithm 2 correspond to the \textcolor{black}{single ground–aerial user pair} and multi‑user schemes, respectively.
\begin{figure}
	\centering
	\includegraphics[width=0.45\textwidth, trim = 10 2 2 0,clip]{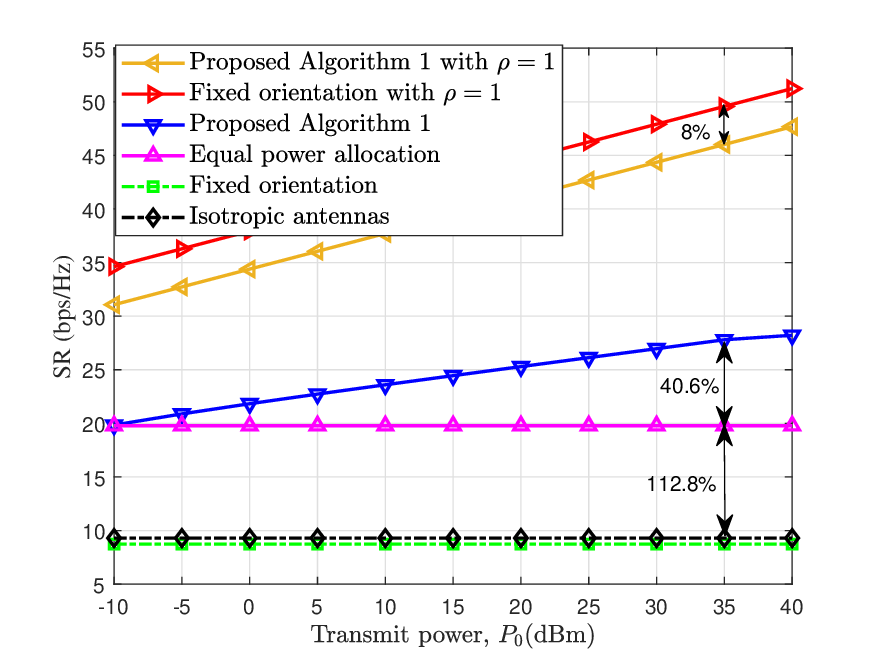}\\
	\caption{SR versus the BS transmit power $P_0$ in the \textcolor{black}{single ground–aerial user pair case.}}\label{fig:2}
\end{figure}
\begin{figure}
	\centering
	\includegraphics[width=0.45\textwidth, trim = 10 2 2 0,clip]{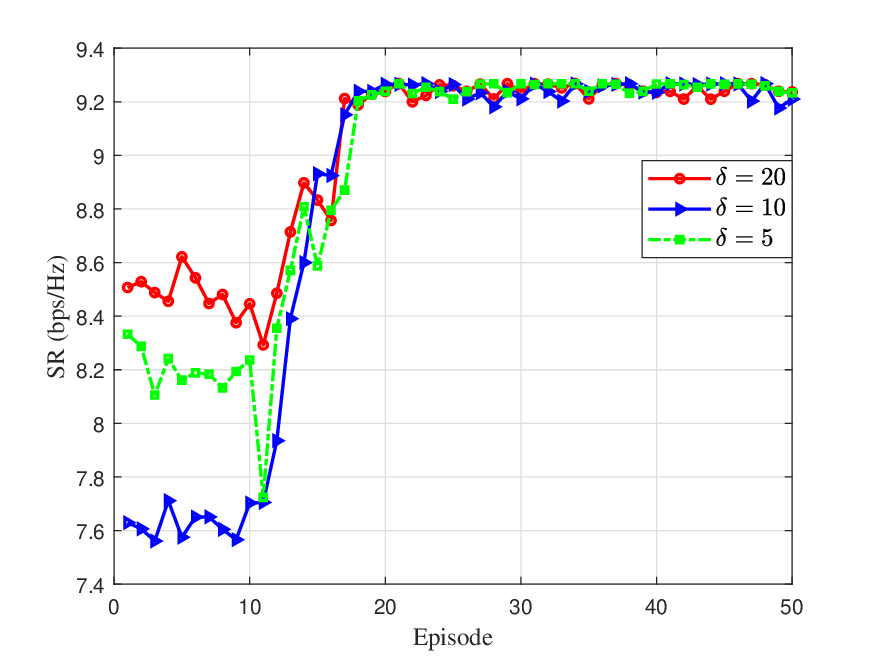}\\
	\caption{ Convergence behavior of the DSAC-T-based algorithm.}\label{fig:3}
\end{figure}
\begin{table}[t]\color{black}
	\centering\renewcommand{\arraystretch}{1.2}
	\setlength{\tabcolsep}{20pt}
	\caption{Detailed Hyperparameters.}\label{lab1}
	\begin{tabular}{lc}
		\hline
		Hyperparameters& Value \\
		\hline		
		Maximum steps per episode&100\\
		Replay buffer size&100000\\
		Sample batch size&64\\			
		Hidden layer dimension&256\\
		Activation function&ReLU\\	
		Discount factor & 0.99\\
		Actor learning rate & 3e-4 \\
		Critic learning rate & 3e-4\\
		Soft update coefficient&0.005\\
		Temperature parameter&0.01\\
		Value distribution range&(0, 30)\\
		Adaptive bound range&[0.5, 2.0]\\
		\hline
	\end{tabular}
\end{table}

First, the SR performance in the \textcolor{black}{single ground–aerial user pair case} is presented. Fig. \ref{fig:2} plots the SR performance versus the transmit power corresponding to the BS. $\rho=1$ indicates that symbols are transmitted only to the aerial Bob, while no signal transmission is performed between FPAAs. For Algorithm \ref{alg1}, by optimizing the RA orientations, serving only the aerial Bob achieves an 8\% improvement in SR performance. When information transmission to a ground Bob is present, the SR performance is degraded by approximately 43.8\%, where inter‑array interference is a significant factor. Under high inter‑array interference, the achievable SR of the benchmark schemes with equal power allocation, fixed orientation, and isotropic antennas remain nearly constant as the transmit power increases. By employing power allocation, the limitation that increasing transmit power does not improve SR performance can be overcome. Compared to the equal power allocation design, a 40.6\% performance gain can be achieved at 35 dBm. Compared with the isotropic antenna benchmark, the equal power allocation scheme achieves a 112.8\% improvement in SR performance, demonstrating the advantage of employing RAs.

\begin{figure}
	\centering
	\includegraphics[width=0.45\textwidth, trim = 10 2 2 0,clip]{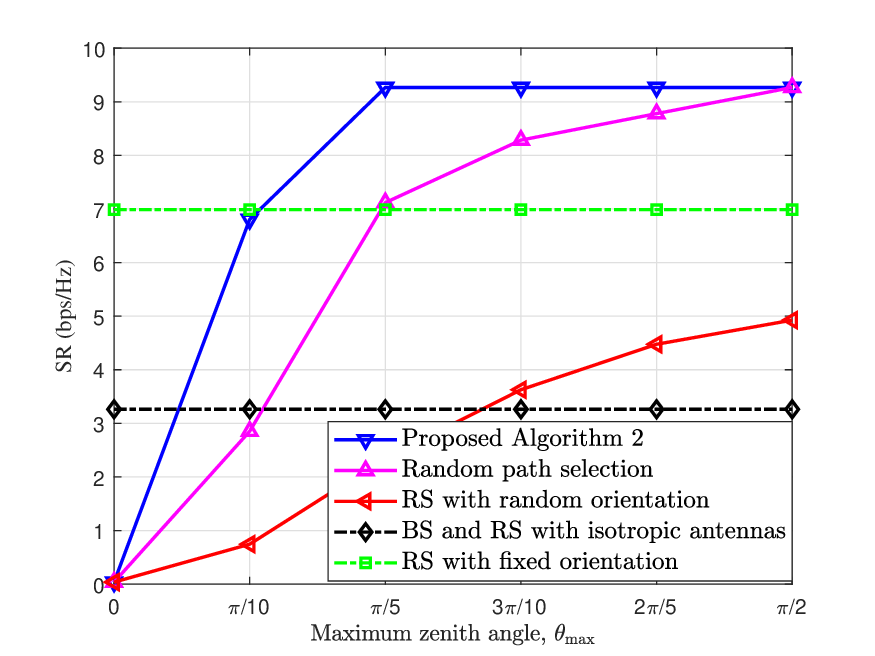}\\
	\caption{SR versus the maximum zenith angle.}\label{fig:4}
\end{figure}

Fig. \ref{fig:3} illustrates the convergence curves under different penalty factors. It can be observed that the DSAC‑T‑based algorithm achieves convergence. The relationship between SR performance and the maximum adjustable zenith angle $\theta_{\max}$ is illustrated in Fig. \ref{fig:4}. As the maximum zenith angle increases, the beam design gains more flexibility to balance the useful signal and interference, thereby improving the SR performance. When the maximum zenith angle is greater than or equal to $\frac{\pi}{5}$, the achievable SR performance of the proposed Algorithm 2 approaches the transmission rate from the BS to the RS. With the random path selection scheme, an SR performance comparable to that of Algorithm 2 can be achieved at a maximum zenith angle of $\frac{\pi}{2}$, demonstrating the importance of path selection optimization. It is noteworthy that, with a sufficiently adjustable zenith angle range, the proposed Algorithm 2 achieves approximately a twofold improvement in SR performance compared to the isotropic antenna benchmark. In the multi‑user scenario, the SR gain of the proposed scheme is particularly significant, as the RAs can effectively reduce multi‑user interference. With random orientations, Bobs may receive weak signals, making it difficult to improve the SR performance. It can be observed that when the maximum zenith angle is less than $\pi/5$, the average SR performance is lower than that of the isotropic antenna benchmark. Fortunately, when the RA orientations are randomly chosen and suboptimal, a low‑cost RIS communication path can be used as an alternative, thereby enhancing the SR performance.
\begin{figure}
	\centering
	\includegraphics[width=0.45\textwidth, trim = 10 2 2 0,clip]{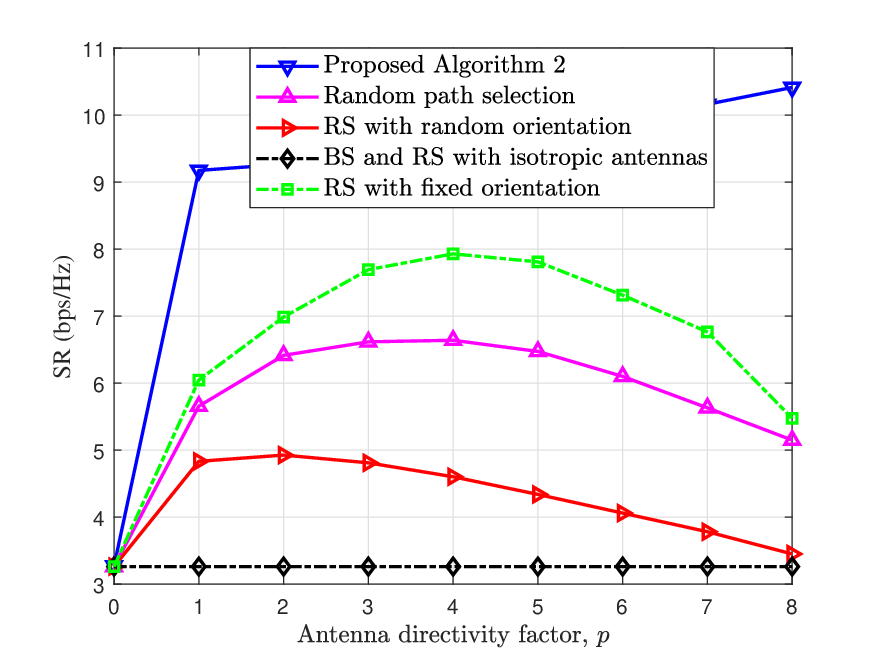}\\
	\caption{SR versus the antenna directivity factor.}\label{fig:5}
\end{figure}

The SR versus the antenna directivity factor is illustrated in Fig. \ref{fig:5}. As the antenna directivity factor increases, higher gain is produced in the main lobe direction, and a narrower main lobe is generated. The achievable SR performance of the proposed Algorithm 2 is gradually improved. Under the other baseline schemes, a high directivity factor is not conducive to achieving high SR performance. Since the rotatable antennas are oriented toward a single target, it becomes difficult to effectively mitigate multi‑user interference when the angular separation among users is small, resulting in a low worst‑case SR for the fixed‑orientation baseline. When $p>4$, randomly selecting communication paths for ground users is not beneficial for improving the SR. To guarantee the QoS for ground Bobs, more transmit power is allocated to them. For 
$p<8$, both the proposed Algorithm 2 and the other baseline schemes outperform the isotropic antenna benchmark.

\begin{figure}
	\centering
	\includegraphics[width=0.45\textwidth, trim = 10 2 2 0,clip]{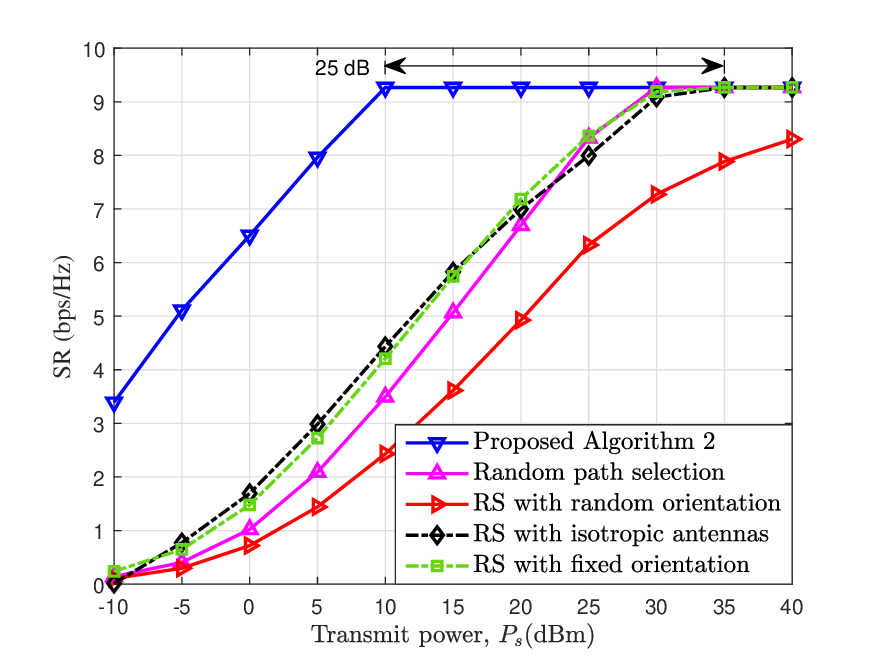}\\
	\caption{\color{black}SR versus the RS transmit power $P_s$ in the multi-user scenario.}\label{fig:6}
\end{figure}
\begin{figure}
	\centering
	\includegraphics[width=0.45\textwidth, trim = 10 2 2 0,clip]{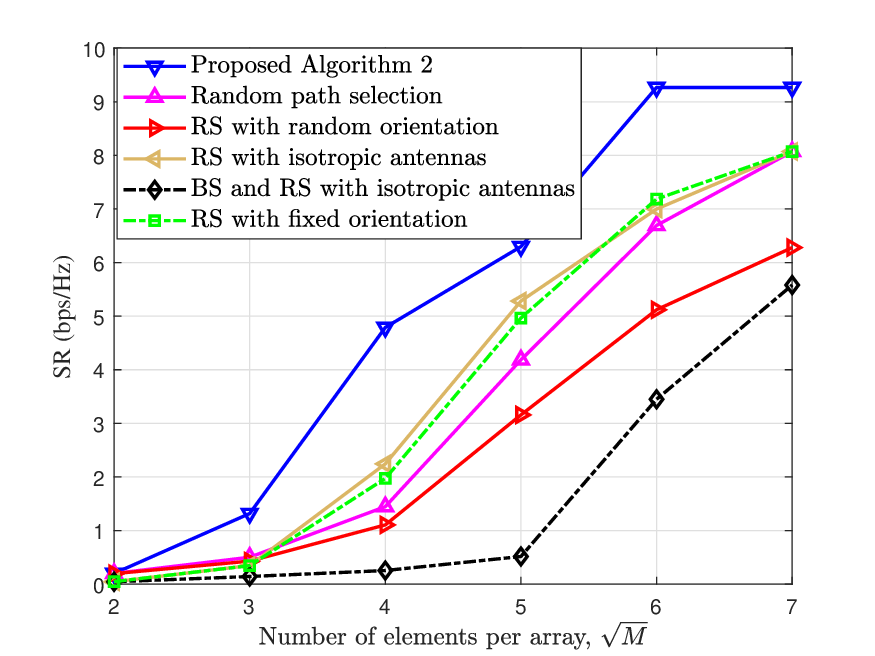}\\
	\caption{SR versus the number of elements per array.}\label{fig:7}
\end{figure}
Fig. \ref{fig:6} illustrates the SR performance versus the transmit power $P_s$. As the transmit power at the RS increases, the SR performance is improved. At $P_s=10$ dBm, the achievable SR of the proposed Algorithm 2 is comparable to the transmission rate from the BS to the RS. The random path selection, RS with fixed orientation, and RS with isotropic antennas baselines reach the RS transmission rate at 30 dBm, 35 dBm, and 35 dBm, respectively. \textcolor{black}{Compared with the RS with isotropic antennas baseline, the proposed Algorithm 2 reduces the required RS transmit power from approximately 35 dBm to 10 dBm for achieving a comparable SR, corresponding to a 25-dB reduction. In the linear power scale, this corresponds to a power saving of approximately 99.7\%.} As the number of antennas per array increases, the SR performance is enhanced, as illustrated in Fig. \ref{fig:7}.  \textcolor{black}{For the proposed Algorithm \ref{alg2}, with $\sqrt{M}=4$ as the reference, the baseline where both the BS and RS are equipped with isotropic antennas achieves comparable SR performance for $\sqrt{M}$ between 6 and 7. The baseline where the RS is equipped with isotropic antennas achieves comparable performance when $\sqrt{M}=5$.} Compared to the benchmark with isotropic antennas at both the BS and RS, the proposed Algorithm 2 \textcolor{black}{can achieve} approximately 55\% antenna savings per array. Compared to the baseline with isotropic antennas only at the RS, Algorithm 2 achieves about 36\% antenna savings per array. Owing to the limited transmission rate at the relay, deploying isotropic antennas at the base station leads to a lower SR performance gain.

\begin{figure}
	\centering
	\includegraphics[width=0.45\textwidth, trim = 10 2 2 0,clip]{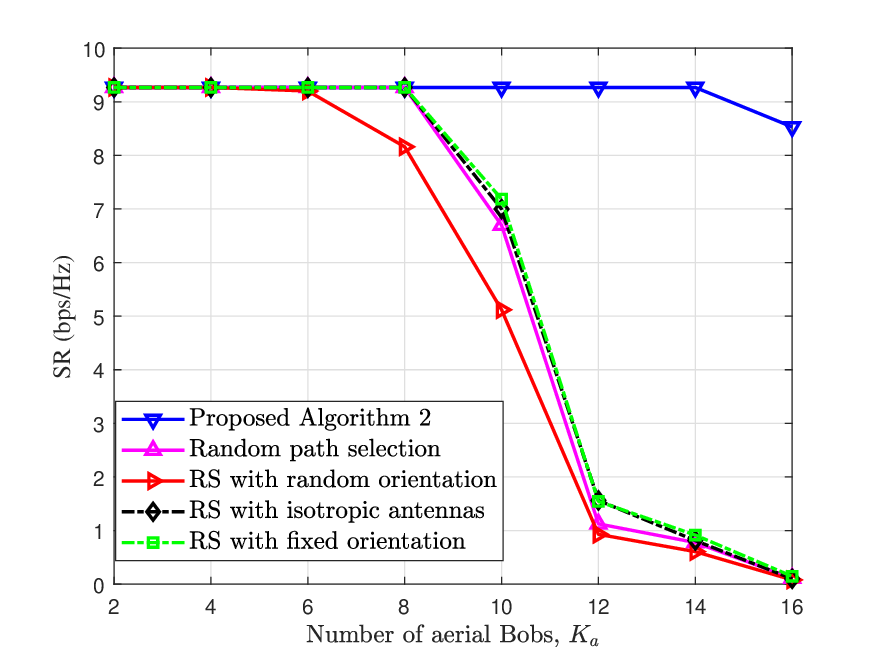}\\
	\caption{SR versus the number of aerial Bobs.}\label{fig:8}
\end{figure}

\begin{figure}
	\centering
	\includegraphics[width=0.45\textwidth, trim = 10 2 2 0,clip]{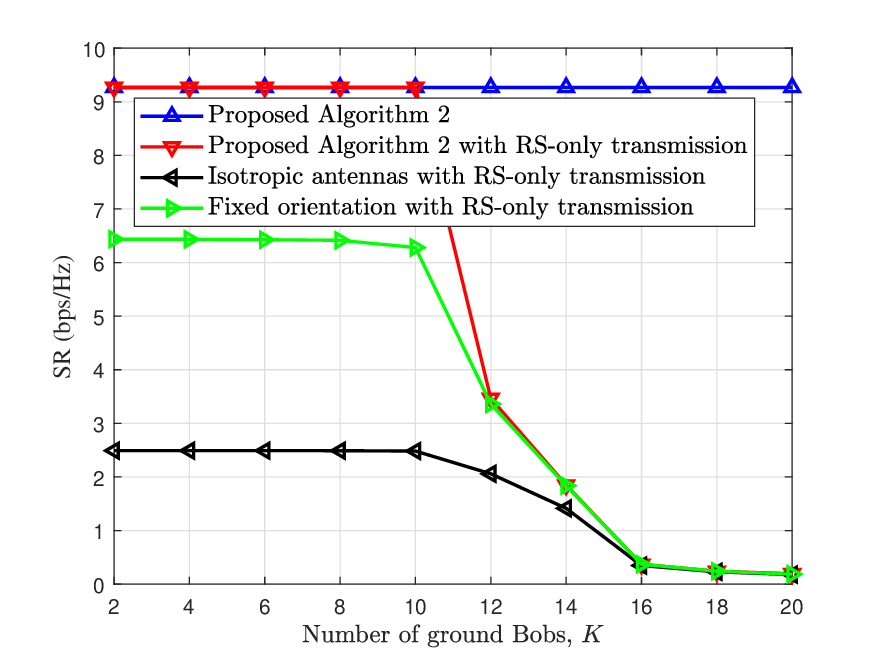}\\
	\caption{SR versus the number of ground Bobs.}\label{fig:9}
\end{figure}

The achievable SR performance under different numbers of aerial and ground Bobs is shown in Figs. \ref{fig:8} and \ref{fig:9}. When the number of aerial users is less than or equal to 14, the proposed Algorithm 2 maintains high SR performance. At $K_a=14$, the other benchmark schemes achieve only about 1 bps/Hz in SR. The proposed scheme can serve more aerial Bobs while still maintaining high SR performance. Fig. \ref{fig:9} further compares the achievable SR performance under RS‑only transmission. When the number of ground Bobs exceeds 10, the SR performance of the benchmark schemes degrades significantly. This is because, with limited transmit power, communication solely through the RS cannot effectively reduce inter‑user interference. It can be observed that the proposed Algorithm 2 maintains high SR performance. \textcolor{black}{In conventional RIS-assisted schemes, the same user may receive the combined gain from both the relay and the RIS, but may also face increased interference and synchronization challenges. The proposed scheme adopts a path selection strategy for ground Bobs, which assigns users to one of the two paths according to their instantaneous transmission conditions, thereby reducing the number of concurrently active links and alleviating the relay load.} Through path selection, inter‑user interference can be effectively mitigated, and more power can be saved for beamforming design to prevent information leakage to Eve.

\begin{figure}
	\centering
	\includegraphics[width=0.45\textwidth, trim = 10 2 2 0,clip]{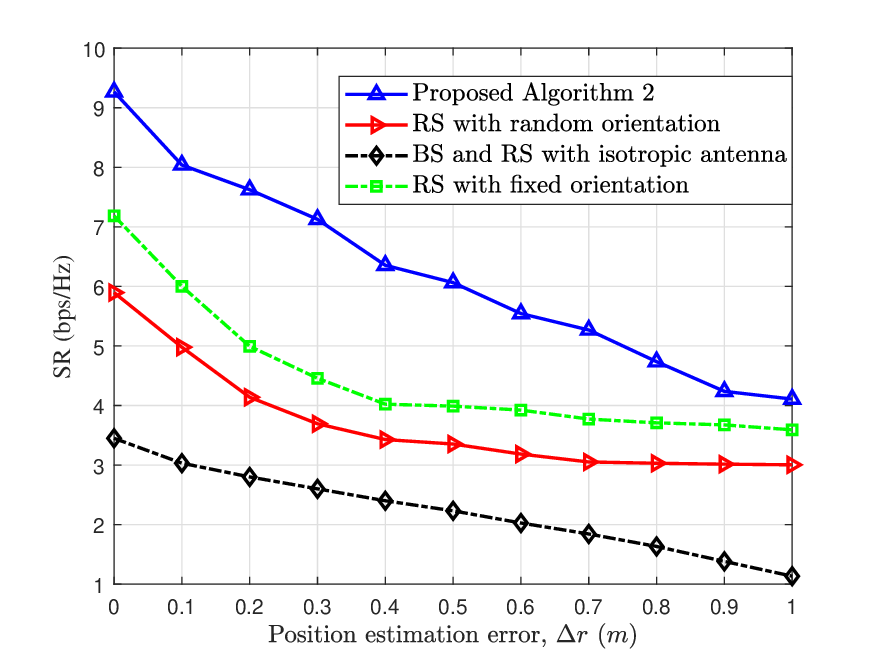}\\
	\caption{\textcolor{black}{SR versus the position estimation error.}}\label{fig:10}
\end{figure}
\textcolor{black}{To further evaluate the robustness of the proposed scheme under imperfect CSI, Fig. \ref{fig:10} illustrates the SR performance versus the position estimation error $\Delta r$. Here, $\Delta r$ represents the uncertainty between the estimated and actual positions along the $x$, $y$, and $z$ axes. Specifically, the variations in $x$, $y$, and $z$ are chosen within the range $[-\Delta r, \Delta r]$. Since the considered LoS channel coefficients depend on the relative positions between the transceivers, position estimation errors result in mismatches between the estimated and actual channel responses. It can be observed that the achievable SR of all schemes gradually decreases as the position estimation error increases. This is because the channel mismatch caused by inaccurate position information degrades both the beamforming design and the orientation optimization of the RAs. Nevertheless, the proposed Algorithm 2 consistently achieves the highest SR over the considered error range. Although its performance advantage gradually decreases for larger estimation errors, it still outperforms the benchmark schemes, demonstrating that the proposed method is robust to moderate CSI uncertainty caused by position estimation errors.}

\section{Conclusion}\label{sec:6}
This paper has investigated the security of a relay network with hybrid fixed‑position and rotatable antenna arrays. The integration of rotatable antenna arrays into existing base stations and relays has been considered. Moreover, a reconfigurable intelligent surface has been shown to provide additional controllable communication paths for ground users, which can be installed on building facades. To maximize the worst‑case SR of the system, both \textcolor{black}{single ground–aerial user pair} and multi‑user scenarios have been studied. A leakage‑based scheme and a learning‑based algorithm have been respectively developed to jointly optimize the beamforming vectors, power allocation factors, and path selection matrix, aiming to maximize the SR performance of aerial users under resource constraints. Simulation results have validated the effectiveness of the proposed algorithms. Through path selection, rotatable antenna orientation optimization, and power allocation, the SR performance is further enhanced compared to conventional relay networks.
\ifCLASSOPTIONcaptionsoff
\newpage
\fi

\renewcommand\refname{References}
\bibliographystyle{IEEEtran}
\bibliography{mybib}

\newpage

\vfill

\end{document}